\documentstyle[ltwol,epsf,psfig,colordvi]{article}

\def\be{\begin{equation}}
\def\ee{\end{equation}}
\def\bea{\begin{eqnarray}}
\def\eea{\end{eqnarray}}
\newcommand{\pbar}{\overline{p}}

\begin{document}

\pssilent

\newcommand{\bdm}{\begin{displaymath}}
\newcommand{\edm}{\end{displaymath}}
\newcommand{\beqa}{\begin{eqnarray*}}
\newcommand{\eeqa}{\end{eqnarray*}}
\title{Precision Calibration of the NuTeV Calorimeter}

\author{
 D.~A.~HARRIS$^{7}$, J.~YU$^{3}$, 
 T.~ADAMS$^{4}$, A.~ALTON$^{4}$, S.~AVVAKUMOV$^{7}$, L.~de~BARBARO$^{5}$, 
 P.~de~BARBARO$^{7}$, R.~H.~BERNSTEIN$^{3}$,
 A.~BODEK$^{7}$, T.~BOLTON$^{4}$, J.~BRAU$^{6}$, D.~BUCHHOLZ$^{5}$,
 H.~BUDD$^{7}$, L.~BUGEL$^{3}$, J.~CONRAD$^{2}$, R.~B.~DRUCKER$^{6}$, 
 J.~FORMAGGIO$^{2}$, R.~FREY$^{6}$, J.~GOLDMAN$^{4}$,
 M.~GONCHAROV$^{4}$, R.~A.~JOHNSON$^{1}$, J.~H.~KIM$^{2}$, S.~KOUTSOLIOTAS$^{2}$,
 G.~KRISHNASWAMI$^{7}$,  M.~J.~LAMM$^{3}$, W.~MARSH$^{3}$, 
 D.~MASON$^{6}$, C.~McNULTY$^{2}$, K.~S.~McFARLAND$^{3,7}$, 
 D.~NAPLES$^{4}$, P.~NIENABER$^{3}$, A.~ROMOSAN$^{2}$,  W.~K.~SAKUMOTO$^{7}$,
 H.~SCHELLMAN$^{5}$, M.~H.~SHAEVITZ$^{2}$, P.~SPENTZOURIS$^{2,3}$, 
 E.~G.~STERN$^{2}$, B.~TAMMINGA$^{2}$, A.~VAITAITIS$^{2}$, M.~VAKILI$^{1}$, 
 E.~VAN ARK$^{7,5}$, V.~WU$^{1}$,  U.~K.~YANG$^{7}$,  
 and G.~P.~ZELLER$^{5}$\\

\begin{center}
(The NuTeV Collaboration)
\end{center}

\vspace{-2.5in} 
\begin{flushright}
UR-1561 \\
FERMILAB-PUB-99-024-E \\  
\end{flushright} 
\vspace{2.15in} 
}
\address{
 $^1$University of Cincinnati, Cincinnati, OH 45221 \\            
 $^2$Columbia University, New York, NY 10027 \\                   
 $^3$Fermi National Accelerator Laboratory, Batavia, IL 60510 \\  
 $^4$Kansas State University, Manhattan, KS 66506 \\              
 $^5$Northwestern University, Evanston, IL 60208 \\               
 $^6$University of Oregon, Eugene, OR 97403 \\                    
 $^7$University of Rochester, Rochester, NY 14627 \\              
}

\twocolumn[\maketitle
\abstracts{
NuTeV is a neutrino-nucleon deep-inelastic scattering 
    experiment at Fermilab.
The detector consists of an iron-scintillator sampling calorimeter
	interspersed
    with drift chambers, followed by a muon toroidal spectrometer.
We present determinations of response and resolution functions
	of the NuTeV calorimeter for electrons, 
	hadrons, and muons over an energy range of 4.8 to 190~GeV.  
The absolute hadronic energy scale is determined to an accuracy of 
0.43\% . 
We compare our measurements to predictions from calorimeter 
theory and GEANT3 simulations. 
}]
%
\section{Introduction}
The increased intensity of the Fermilab Tevatron fixed-target
	program has made it possible to qualitatively improve neutrino
	deep-inelastic scattering experiments. 
Deep-inelastic neutrino scattering probes both the electroweak 
	and strong forces in unique ways which are
	both competitive and complementary to other measurements at 
	hadron and electron colliders. 
For these reasons, it is important to continue improving the 
	precision of measurements with neutrino beams.
NuTeV (Fermilab Experiment 815) is designed to exploit the intensity
	capabilities at Fermilab using a new neutrino beam, an 
	upgraded neutrino detector, and a continuous test beam 
	calibration system. 

\begin{figure}[tbph]
\begin{center}
\centerline{\psfig{figure=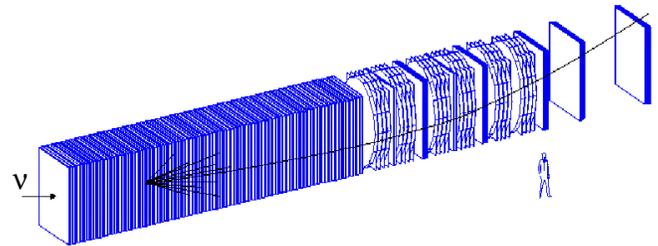,width=0.5\textwidth}}
\caption{The NuTeV neutrino detector showing the target calorimeter followed 
by the downstream muon spectrometer.}
\label{fig:detector} 
\end{center}
\vspace{-.4in} 
\end{figure}

The new neutrino beam uses a sign-selected, quadrupole train 
	(SSQT)~\cite{ssqt} to produce a high-intensity,
	 ultra-pure beam of either neutrinos or antineutrinos. 
For neutrino detection, the experiment uses an upgraded
	version of the CCFR detector~\cite{elsewhere}, shown in 
Figure~\ref{fig:detector}, with new scintillation oil and photomultiplier 
	tubes combined with refurbished drift chambers. 
In the detector, neutrino interactions produce a hadronic shower from the
	outgoing struck quark whose energy is measured in the 
	target-calorimeter and, for charged current events, an 
	associated outgoing muon whose angle (momentum) is measured 
	in the target-calorimeter (downstream muon spectrometer). 

The NuTeV data run took place during the Fermilab 1996-1997 fixed
	target run.  
The experiment recorded over three million neutrino and
	antineutrino interactions. 
Two of the physics goals of NuTeV are a precise measurement of the weak
	mixing angle and measurement of structure functions and 
	the strong coupling constant from QCD scaling violations.  
Both of these results are dependent upon a detailed understanding of 
	the response of the target-calorimeter.  
A previous experiment using this calorimeter, CCFR, determined the 
	calorimeter energy scale to an uncertainty of approximately $1\%$.  
CCFR measures~\cite{CCFR-wma} the weak mixing angle to be 
$\sin^2\theta_W~=~0.2236~\pm0.0019{\rm\textstyle (stat)}~\pm0.0019{\rm\textstyle (syst)}~
\pm0.0030{\rm\textstyle (model)}$.  
NuTeV aims for a total precision of better than $0.002$ on $\sin^2\theta_W$,
 	primarily by changing the measurement technique to reduce 
	model uncertainties.
However, in CCFR the experimental systematic uncertainty due to 
	calorimeter response is $\pm 0.0011$ and the NuTeV technique 
	is considerably more sensitive to energy calibrations.  
The reduced theoretical uncertainties make an improved calibration 
	essential for the success of this measurement.  

In the case of the measurement of the strong coupling constant, the 
	systematic uncertainty on the QCD scale parameter,
	$\Lambda _{\overline{MS}}$, from calibration effects is 
	at the 50-100~MeV level previously\cite{Seligman-alphas}, 
	which is the largest single experimental source of uncertainty 
	in the measurement.  
In NuTeV, this uncertainty would be reduced by a factor of three
 	by an absolute calibration of $0.3\%$ uncertainty.

For the measurements listed above, the absolute response and 
	resolution of the hadronic shower energy
	measurement are crucially important.  
For this reason, precision detector calibration and response 
	determination is a key component of the NuTeV program.  
This is done using several data sets: the actual neutrino events, 
	neutrino-induced muons from upstream shielding, and 
	calibration beam data.
Throughout the data run, the calibration beam operates
	continuously and provides momentum tagged electrons, muons,
	and hadrons with energies between 4.8 and 190 GeV.  
A precision spectrometer provides an event-by-event momentum
	determination with resolution better than $0.3\%$ and a
	combination of a \u{C}erenkov counter and a TRD are 
	used to determine the particle type for each event.

This article describes the various techniques and studies
	that lead to the precision calibration of this calorimeter.  
First, the detector and electronics calibrations 
	using neutrino-induced events are described, followed by 
	the test of these techniques and resolution studies 
	using the calibration beam data.  
This article also demonstrates which
	aspects of the detector response are accurately modeled by 
	GEANT and other software simulation packages.


%
\section{The NuTeV Calorimeter}\label{ss:detector}
\begin{figure}[tbph]
\begin{center}
\centerline{\psfig{figure=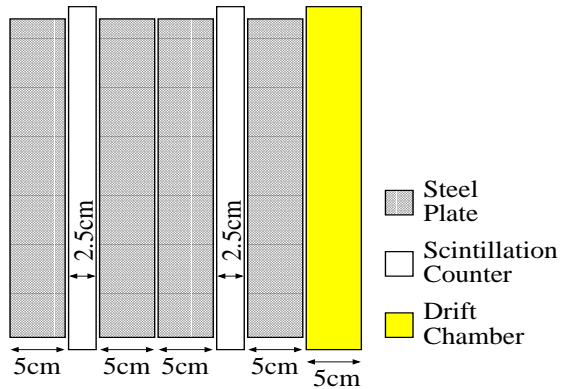,width=0.4\textwidth,height=2.in}}
\caption{Geometry of one unit of the calorimeter.  This unit is repeated
42 times to make up the entire calorimeter.  One unit of the
calorimeter consists
of a scintillation counter sandwiched between two steel plates.}  
\label{fig:target} 
\end{center}
\end{figure} 
The NuTeV calorimeter consists of 168 plates of steel measuring 
	$3~{\rm m~(H)} \times 3~{\rm m~(W)} \times~{\rm 5.1~cm~(L)}$, 
	interspersed with 84 scintillation counters of dimension 
	$3~{\rm m~(H)} \times 3~{\rm m~(W)} \times~{\rm 2.5~cm~(L)}$ and 
	42 drift chambers.  
There are two plates of steel between every two consecutive scintillation
	counters, and one drift chamber between every other set of counters.  
One unit counter consists of a scintillation counter and two steel plates
	surrounding the scintillator.
Therefore one unit calorimeter layer consists of 
	two counters and a drift chamber.
This configuration leads to a detector with 10.35~cm of steel between counters
	and 20.7~cm of steel between drift chambers.
The geometry of one unit of the calorimeter is shown in 
	Figure~\ref{fig:target} 
	and this unit is repeated 42 times to make up the 
	entire calorimeter.   
Table~\ref{tab:target} summarizes the materials 
and their longitudinal sizes in units of cm, 
	radiation length, and interaction length, for one unit of 
	the calorimeter's longitudinal layer.
\begin{table}[bp]
\caption{Composition in interaction length and radiation length of one 
unit of the NuTeV calorimeter.  
This unit is repeated 42 times to make up the entire calorimeter.}  
\label{tab:target} 
\begin{center}
\begin{tabular}{lrrr} 
\hline
Component & & Length &\\
& cm & $X_0$ & $\lambda_I$ \\ 
\hline
4 Steel Plates & 20.7 & 11.75 & 1.24 \\ \hline 
2 Scint. Counters & 13.0 & 0.51 & 0.16 \\ \hline 
1 Drift Chamber & 3.7 & 0.17 & 0.03 \\ \hline\hline
Total & 37.4 & 12.43 & 1.43 \\ \hline 
\end{tabular}
\end{center}
\end{table}

\begin{figure}[tbph]
\begin{center}
\centerline{\psfig{figure=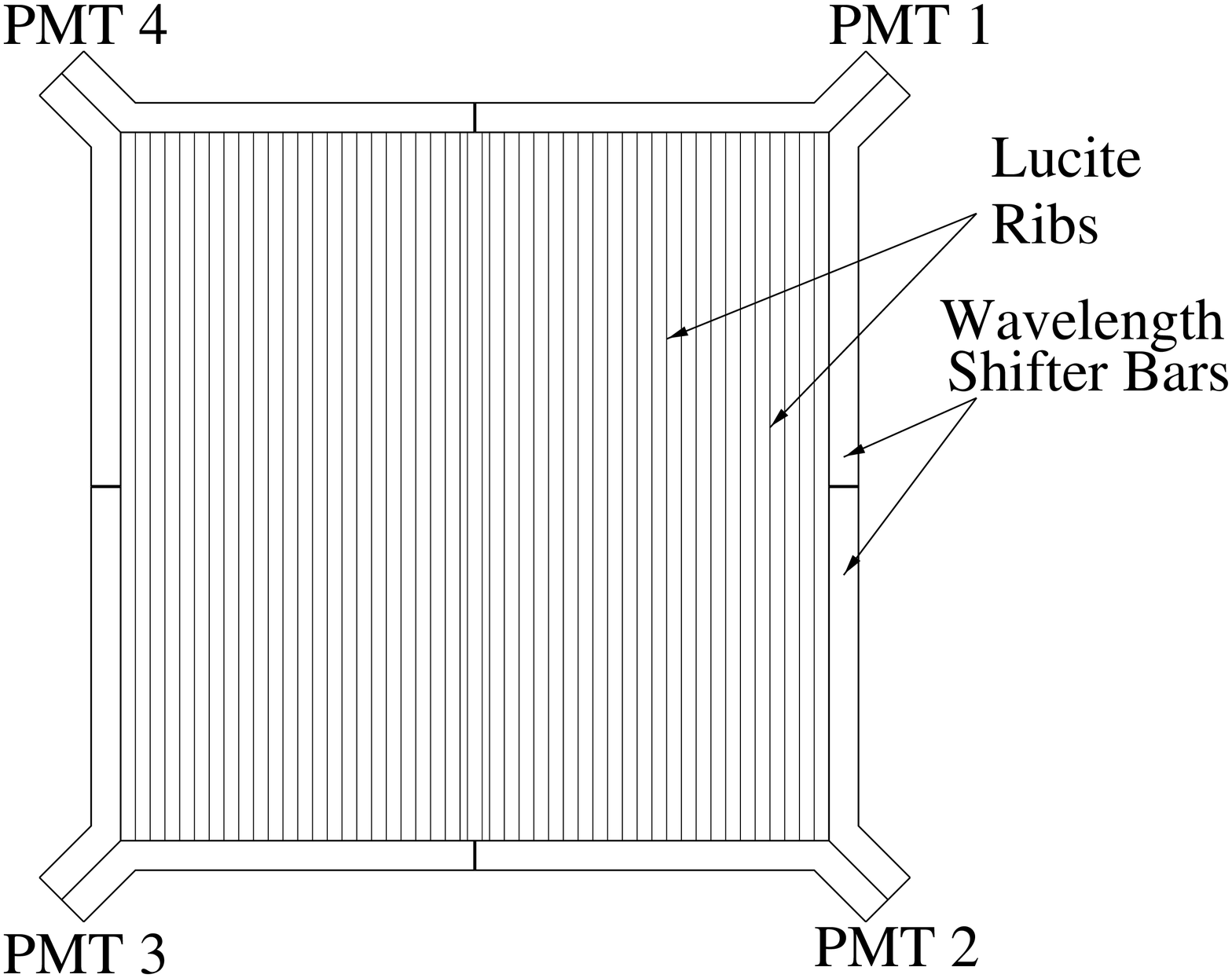,width=3.5in,height=3.in}}
\end{center}
\vspace{-.3in} 
\caption[]{A schematic drawing of a NuTeV scintillation counter.}
\label{fig:counter-schematics}
\end{figure}

Figure~\ref{fig:counter-schematics} shows a schematic diagram of a NuTeV
	scintillation counter.
The scintillation counters are lucite boxes filled with 
	Bicron 517L scintillator oil.  
The counters have 3~mm vertical lucite ribs spaced by 2.5-5.1~cm,
	depending on the lateral position of the ribs, 
and designed for structural support.  
Since these ribs do not scintillate, the counters are staggered so that 
the ribs 
are not aligned on the transverse plane throughout 
	the length of the calorimeter.  

Each counter is surrounded by 8 wavelength-shifter bars, doped with
	green BBQ fluor, and is read out in 
	four corners by photomultiplier tubes (PMT's), mounted one on
	each corner.
The PMT's are 10-stage Hamamatsu R2154 phototubes with a 
	green-extended photocathode, with gains set to about $10^6$.  

There is an air joint between the wavelength-shifter bars.
The joints between the wavelength-shifter bars and the phototubes 
	have 3~mm thick clear silicon jelly cookies for better 
	optical and mechanical connections, 
as well as for PMT
	window protection. 
The cookies are made of Dow Corning Sylgard(R) 182 silicon elastomer and 
	182 curing agent.
Given this particular geometry and readout scheme, NuTeV observes muon 
	signal distributions which are consistent with, on average, 
	30 photoelectrons for muons traversing the center of a counter.  
For muons closer to the edge of the counter, where light collection is 
	more efficient, the number is higher.  
%

%
\section{Calorimeter Readout Electronics}\label{ss:electronics}
The design of the readout electronics is dictated by the requirement to
	accommodate a very large dynamic range of the signal 
	using 11-bit analog-to-digital converters (ADC's). 
Minimum ionizing particle energy loss (MIP) in the calorimeter is 
	approximately $0.15$~GeV per unit counter while neutrino 
	interaction induced hadronic showers could deposit up to 
	$100$~GeV into a single unit counter. 
Note that the actual energy deposit of a minimum ionizing particle 
	in a single scintillation counter is approximately 4~MeV.

\begin{figure}[ptb]
\begin{center}
\psfig{file=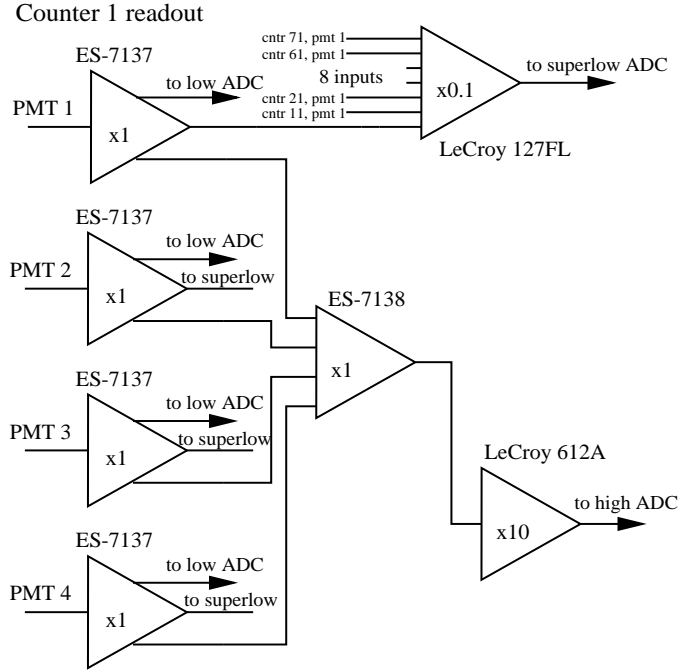,clip=,height=3.5in,width=3.5in}
\end{center}
\vspace{-.3in} 
\caption[]{ A schematic diagram of the NuTeV calorimeter readout electronics
	of a counter.}
\label{fig:electronics}
\end{figure}

Figure~\ref{fig:electronics} shows a schematic diagram of the 
	readout electronics system.
The readout electronic channels consist of the following three separate 
	gains, to measure energies in a wide dynamic range:
\begin{itemize}
\item{$HIGH$ is the signal formed by a sum of 
	signals from each of the four PMT's of a given counter
	(fan-in ES-7138)~\cite{es7138} and $\times10$ amplification of 
	the summed signal by the linear amplifier LeCroy 612A. 
The amplified signal is then digitized by a LeCroy 4300~\cite{lecroy-4300fera} 
	Fast Encoding and Readout ADC (FERA).} 
\item{
$LOW$ is the signal from each of the 4 PMT's directly digitized
	by LeCroy 4300 FERA.}
\item{$SUPERLOW$ signals are the digitized sums of 8 PMT signals which
	come from 8 different counters, each separated by 10 counters. 
	Each PMT signal is attenuated by 1/10 
	(fan-in LeCroy 127FL~\cite{lecroy-127fl}).}
\end{itemize} 

A typical minimum ionizing particle signal produces 80~ADC counts in 
	the $HIGH$ channel, 2~ADC counts in each $LOW$ channel, and 
	$\sim0$~ADC counts in each $SUPERLOW$ channel. 
The $LOW$ and $SUPERLOW$ channels are calibrated with respect to 
	the $HIGH$ channels.

Hadronic showers from neutrino interactions deposit up to 600~GeV 
	in the calorimeter (with maximum energy deposition in a 
	single counter of about $100$~GeV).
A typical hadronic shower in the calorimeter saturates the $HIGH$ 
	channels and leaves a signal of a few hundred ADC counts
	in the $LOW$ channels. 
Thus the $LOW$ channels are used to measure the shower energy. 
In a very small fraction of the time, one of the four $LOW$ channels 
	of the counter is saturated when the transverse position of 
	a neutrino interaction is close to one of the PMT's.
In these cases, the attenuated $SUPERLOW$ channel is used.

Calibration of the NuTeV calorimeter
	begins with the calibration of the readout electronics and tying
	the $HIGH$, $LOW$, and $SUPERLOW$ channels into a linear model.

%
\section{Pedestal Subtractions}\label{ss:pedestals}
The best way to determine the pedestal values of the ADC channels well 
	is measuring them under exactly the same conditions as the neutrino
	data.  
To achieve this we use two different methods -- one using a
	specially designed random trigger and the other using ``quiet
	regions'' of the calorimeter during real neutrino events.  
A random trigger is activated throughout the run in all gate types 
	to measure pedestals under the same condition as the trigger 
	of interest.  
The rate of the pedestal trigger is prescaled to provide the necessary
	number of pedestal events -- typically $10$ events per accelerator
	cycle -- without overloading the bandwidth.

The second method uses events where the trigger (T2) is designed for
	neutral current interactions and requires significant energy
	deposition in consecutive counters in the calorimeter.  
For each T2 event, an offline analysis program finds ``quiet
	regions'' in the calorimeter, using the following algorithm:
\begin{enumerate}
\item{Count the number of counters with pulse heights more than 1/4 of a MIP 
	(S-bit ON) and reject the event if this number is larger than 10.}
\item{Select the counters with their own S-bits and the S-bits of
	their 4 closest neighboring counters OFF.}
\item{Check that all three readout ADC channels of the selected counters 
        described in 
	Section~\ref{ss:electronics},
	have energies less than 0.3~MIP.
The cut value is chosen to be much less than 1~MIP
	but much larger than the pedestal; for example 
	in a $HIGH$ ADC channel, one MIP is $\sim$70 ADC counts 
	and a typical pedestal width is $\sim$3 counts, thus the cutoff value 
	of 20 counts is $\sim$7 standard deviations from zero.}
\item{Use the readouts of each ADC channel of the selected counters 
	as the pedestal values.}
\end{enumerate}

The offline analysis procedures use these pedestal events
	to keep a running average for each electronics channel,
	using both these methods.
The two procedures for measuring pedestals agree to within .015 ADC 
	counts in the LOWs, and .02 ADC counts in the HIGHs.  
This pedestal uncertainty would contribute a constant 
	term of 32~MeV to the hadron energy resolution if all pedestal 
	differences were correlated, and a 3.6~MeV width if these pedestal 
	differences were uncorrelated.  
The constant term in the hadron energy resolution is consistent with 
	zero with an error of 110~MeV (see Figure~\ref{fig:hadres}).  
The ``quiet region'' method is used for neutrino data pedestal subtraction.  

Because of the differences in the upstream magnet currents and 
	detector environment, the pedestals during the calibration beam 
	gate and the neutrino beam gate are not necessarily equal.  
In fact, some channels differ by as much as 0.3 ADC counts in the LOWs.  
For the analyses of the calibration beam data, we use the random trigger 
	method to measure the pedestals because the upstream part of 
	the calorimeter, where the calibration hadron beam enters, 
	always has energy deposited in every event, precluding the 
	``quiet region'' method.  
The neutrino data pedestal comparisons ensure 
	that this treatment is completely accurate to the few MeV level.

%
\section{Electronics Cross Calibration}\label{ss:cross-calibration}
Relative calibration of the different channels of
	electronics is needed because the minimum
	ionizing particle signal is measured with the $HIGH$ 
	channels, while the neutrino interaction signal is 
	measured with the $LOW$ and $SUPERLOW$ channels. 
Following the assumption of linearity of all the components of the 
	readout electronics, we assume that the $HIGH$ channel is 
	the linear combination of the 4 $LOW$s
	of the same counter:
\begin{eqnarray}\label{eq:hi-lo-ratio}
HIGH(i) = \sum_{j=1}^{4}R_{j}^{hl}(i)\times LOW_{j}(i),
\end{eqnarray}
where $i$ is the counter index, $R_{j}^{hl}(i)$ is the relative calibration
	constant between the $LOW$ signal of the PMT $j$ of the 
	counter $i$ and the $HIGH$ signal of the counter $i$.

The $SUPERLOW$ is the linear combination of the 8 LOWs, as follows:
\begin{eqnarray}
SUPERLOW(i) = \sum_{k=1}^{8} R^{sl}_{j}(i)\times LOW_{j}(k), 
\end{eqnarray}
where $i$ is the $SUPERLOW$ channel number, $k$ is the counter
	index, $j$ is the fixed PMT index, and  $R^{sl}_{j}(i)$ is
	the relative calibration constant between the $SUPERLOW$ channel $i$
	and the $LOW$ channel of the PMT $j$ of the counter $k$.
The set of calibration coefficients $R^{hl}_{j}(k)$ and $R^{sl}_{j}(k)$
	is calculated for every data-taking run using the 
	least squares method.

%
\section{Counter Gain Position and Time Dependence}\label{ss:gain_cor}
As do all high energy neutrino experiments, NuTeV has an ideal 
	calibration source to track counter gains: muons coming from 
	upstream neutrino interactions in the 
	shielding.
This means that not only are the muons completely correlated in 
	time with the actual neutrino beam, but also illuminate
	the detector in a similar fashion.  
This section describes how the sample of muons traversing the 
	entire length of the detector during the 
	neutrino beam is used to monitor the position and 
	time dependencies of the individual counter gains.  

\begin{figure}[tbp]
\begin{center}
  \epsfxsize=.5\textwidth
  \epsfbox{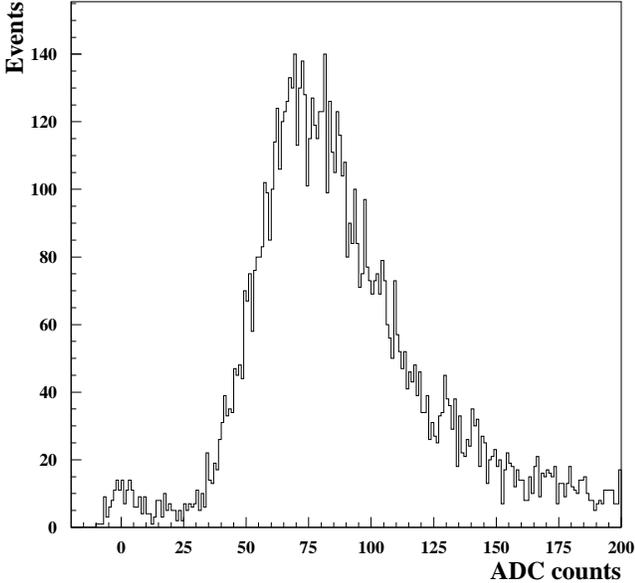} 
\end{center}
\vspace{-.3in} 
\caption[]{Typical energy deposition of muons traversing one scintillator 
	counter in units of ADC counts.}
\label{fig:mupeak} 
\end{figure}

Figure~\ref{fig:mupeak} shows a typical energy deposition profile 
	for muons traversing a particular counter.  
There are, on average, 30 photoelectrons per MIP
	per counter.  
Events with very low pulse heights are from particles that go 
	through the ribs of a counter while the events with large 
	pulse height are from muon bremsstrahlung and $e^{+}e^{-}$ 
	pair production.  
Since pair production increases with increasing muon energy, 
	this energy deposition pattern can be used as an 
        event-by-event muon momentum measurement, as described in
	Reference 7.

In order to characterize the distribution shown in 
	Figure~\ref{fig:mupeak} in a way that is stable with 
	respect to cuts, we use the truncated mean procedure~\cite{auchin}.  
The truncated mean is determined by calculating 
	the mean of the distribution using all events, then taking 
	the mean again but only including the events between 
	0.2 and 2 times the previous mean.
This procedure is iterated several times until the difference between
	the mean of two consecutive iterations is less than 0.1\% of the
	previous mean.
Corrections are made event-by-event for the muon's momentum as well as 
	the angle with respect to the direction perpendicular 
	to the counter.  
This procedure provides a ``mean'' that is insensitive to the
	width and tails of the pulse height distribution.
The truncated mean for 77 GeV muons is defined as 1 MIP.

\subsection{Position Dependence}\label{ss:muon-map}
\begin{figure}[tbp]
\begin{center}
 \epsfxsize=.5\textwidth
  \epsfbox{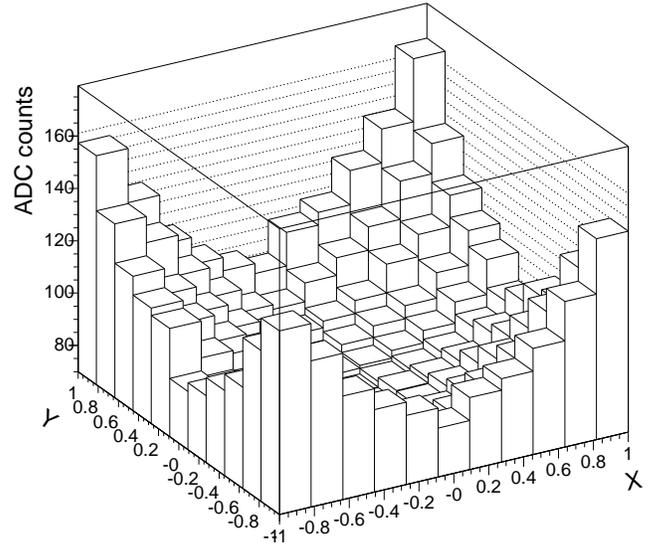}
\end{center}
\vspace{-.3in} 
\caption[]{ Average counter response to muons traversing as a function
of position in the counter.  The coordinates are normalized to the
half width of the counter, 1.5~m, on both axes.}
\label{fig:mumap} 
\end{figure}

We calculate the average position dependence of the truncated means (which 
	will be referred to as the muon response) for each counter,
	by averaging over the entire neutrino run. 
Figure~\ref{fig:mumap} shows that response as a function of position 
	for a typical counter.  
The light collection is largest at the corners of the counters, near the 
	four phototubes, as expected.  
The technique used to track the time dependence
	of the gains alters the position dependence of each 
	counter as a function of time.
This is because the gain for each phototube is determined independently as
 	a function of time and is reflected into the position dependence.  

\subsection{Time Dependence} 
The gain of a single counter at a particular moment in time during the 
	run depends on the gains of the four 
phototubes 
	as well as the gain of the scintillator oil itself.  
To determine the time dependence of the counter gains, we calculate 
	a fractional phototube map, 
which is defined as the fraction of light
	reaching a given phototube as a function of position within 
	an independent counter.  

These fractional phototube maps are measured using high energy neutrino
	interactions where the pulse height is 
\begin{figure}[tbph]
\begin{center}
  \epsfxsize=.5\textwidth
  \epsfbox{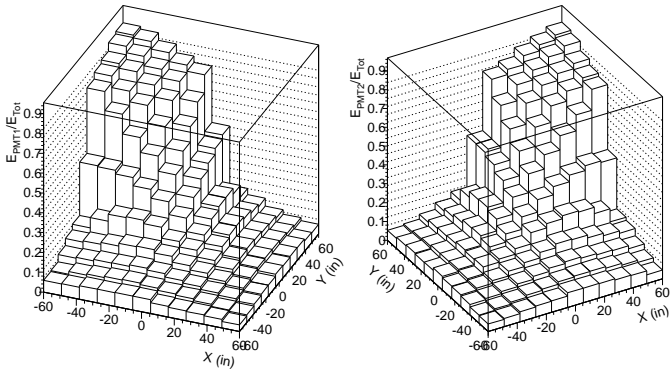}
\end{center}
\caption[]{Fractional energy deposition as a function of position for 
two phototubes in a given counter.}  
\label{fig:pmtmap} 
\end{figure} 
high enough to be seen
	in an individual $LOW$ channel.
Figure~\ref{fig:pmtmap} shows two sample phototube maps. 
	Notice that, as expected, the fractional maps are strongly 
	peaked near the phototubes themselves and drop off sharply 
	where two wavelength-shifter bars meet in the center of the counter.  

We then fit the muon response over a short period of time to a 
	function with four parameters, where the parameters correspond 
	to the gains of the four phototubes.  
So, if $M^0(x,y,i)$ is the run-averaged 
	muon response map for counter $i$ as a 
	function of $x$ and~$y$, and $F_j(x,y,i)$ 
	is the fraction of light reaching phototube $j$ of 
counter~$i$, 
	then the run-averaged muon response for phototube $j$ of 
	counter $i,$ $P_j(x,y,i)$, is simply
\begin{eqnarray}
P_j(x,y,i)=M^0(x,y,i)\times F_j(x,y,i).
\end{eqnarray}
The time-dependent function is then 
\begin{eqnarray}
M(x,y,i,t) = \sum_j g_j(i,t)P_j(x,y,i),
\end{eqnarray}
where $g_j(i,t)$ is the relative gain of phototube $j$ of counter $i$ at 
	time $t$ compared to the average gain over the entire run.  

\begin{figure}[tbph] 
\begin{center}
\centerline{\psfig{figure=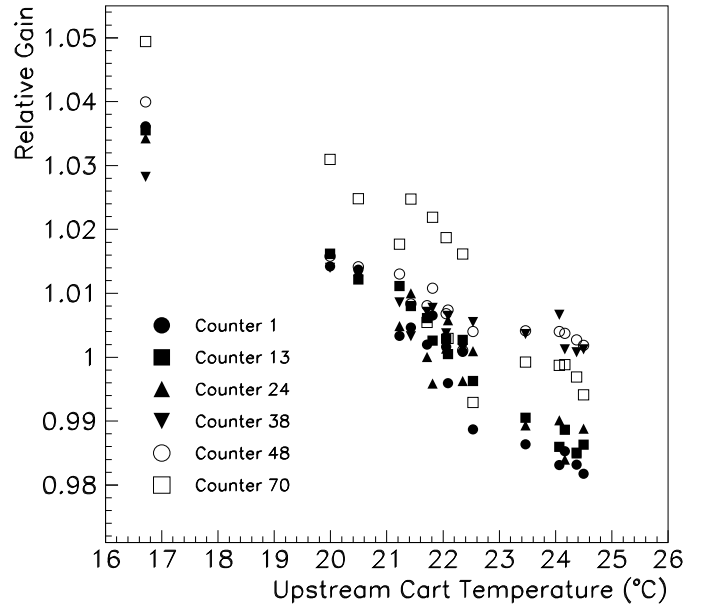,width=.5\textwidth}}
\end{center}
\vspace{-.3in} 
\caption[]{Sum of phototube gain coefficients, normalized to the average
	gain over the entire neutrino run period, as a function of 
	temperature of the most upstream region of the calorimeter.}  
\label{fig:gain_vs_temp} 
\end{figure}

Gains vary by as much as 10\% as a function of time
	and vary, on average, $-0.16\%/^\circ C$ (compared to the value
	$ -0.11\% /^\circ C$ specified by the vendor~\cite{hamamatsu}), 
	as a function of temperature as shown in 
	Figure~\ref{fig:gain_vs_temp}.  
The temperature dependence varies
	from counter to counter by $\pm 0.04\%/^\circ C$.

%
\section{ Hadron Gain Balance} 
\label{ss:hadgainbal} 
The technique described in the previous section 
	determines the gain for each scintillator counter 
	relative to all the others, using muons that traverse 
	the entire detector.  
To set the absolute hadron energy scale of the 
	detector we measure the response of the calorimeter to a 
	monochromatic beam of hadrons incident on the most upstream
	part of the calorimeter (see Section~\ref{sec:hadron} for details).  
Since this hadron beam deposits all of its energy in the most 
	upstream 10-12 counters, the hadron response of  
	only those counters is measured.  
The hadron response measured this way would be usable for the
	entire calorimeter if the hadron response were completely 
	correlated to the muon response.
However, geometric non-uniformities in the calorimeter give rise to  
	relative differences between the hadron and muon response, 
	breaking the correlation.   

The NuTeV calorimeter measures the hadron energy by sampling the
	shower every 10~cm of steel.  
The energy deposited by a hadronic shower in the scintillation 
	counters is only a small fraction
	of the total energy deposited in the detector.  
Therefore, variations in the passive material surrounding 
	each counter affect the average hadron signal sampled 
	in that counter.  
In contrast, the muon signal is only dependent on the variations 
	in the active material.  
Since the relative gain of a counter for hadrons may not be
	completely correlated with the relative gain for muons, 
	setting the hadron energy scale for the first 10--12 
	counters is not sufficient to set the scale for the 
	entire detector.  
In this section we describe the technique used to measure the 
	hadron/muon gain ratio for each counter, using neutrino 
	interactions that occur throughout the entire calorimeter.

Apart from the low interaction rates, neutrinos are a perfect 
	relative hadron calibration source for the entire calorimeter.  
First of all, if the detector is far enough away from the 
	neutrino production target, the energy distribution 
	of neutrinos interacting in the most upstream counter 
	of the calorimeter is the same as that of the neutrinos 
	interacting in the last counter.  
In a charged current $\nu_\mu$ or $\overline\nu_\mu$ 
	interaction, which is the majority of neutrino 
	events that NuTeV sees, both a hadron shower and 
	a muon are produced and deposit energy in the calorimeter.  
As stated previously, the hadron shower deposits most of its 
	energy in the first few counters after the neutrino
	interaction, while the muon deposits a small amount of 
	energy in each counter over many counters, depending on 
	its angle and energy.   

The average measured energy in the calorimeter from neutrino 
	interactions should not depend on where the neutrino 
	interaction occurs, assuming that one always measures 
	the energy by summing over the same number of counters 
	from the event vertex.  
If one sums over the first 10 counters after the event vertex, 
	then the muon contribution to a $70\,{\rm GeV}$ shower is 
	about $3\%$.  
The additional muon energy deposited in the hadronic 
	shower region would reduce the measured effect, but the amount 
	by which the muon's presence changes the measurement is negligible
	compared to the statistical uncertainty on the gains.  

 To determine the hadron/muon gain ratio for each counter, we need 
	a sample of clearly identified neutrino interactions in 
	the calorimeter.  
Events are selected by requiring a final state muon, visible by 
	a minimum energy deposition extending over at least 20 
	counters ($2\,{\rm m}$ of steel equivalent) after the 
	event vertex.  
To remove cosmic ray backgrounds, events are also required to 
	have a reconstructed hadron energy greater than $20\,{\rm GeV}$.  
The hadron energy is determined by summing over the energies of 
	the 10 counters following the event vertex.  
Since this energy cut ultimately depends on the relative gains 
	obtained from the technique, the procedure must be iterated.  
The average hadron energy of the events passing all cuts is about 
	$70\,{\rm GeV}$.  

Events are also required to occur at least 4 counters from the 
	upstream end of the calorimeter, 20 counters before
	the downstream end of the calorimeter, and 
	within $1.27\,{\rm m}$ of the center of the detector.  
These fiducial cuts ensure that the event is not induced by a 
	charged particle entering from the side or front of 
	the detector, and that the hadronic shower in the event 
	is fully contained within the calorimeter.
Because of these fiducial volume cuts, we are unable 
	to use this technique to determine the relative gains 
	of the 15 most downstream or 4 most upstream counters.  
However, the most upstream counter hadron/muon gain ratios 
	are determined using a similar technique, described here, 
	with the calibration beam hadron data.  
The first four counters' gains are set by comparing the calibration 
	beam hadron response of showers starting in the most upstream
	set of four counters with those in the next set of four 
	counters that are immediate downstream of the first set 
	and whose gains have been determined from the neutrino data.  
 
\begin{figure}[tbp] 
\begin{center}
\centerline{\psfig{figure=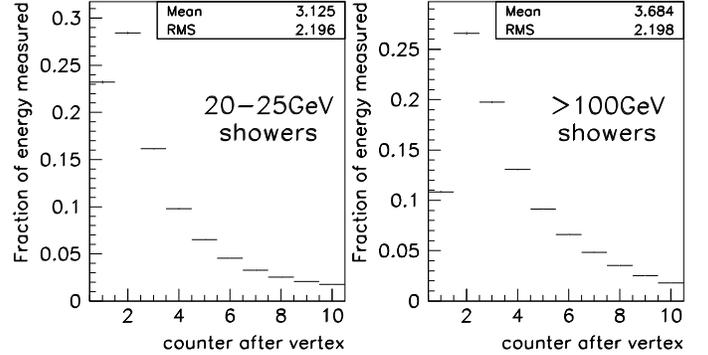,width=.5\textwidth}}
\end{center}
\vspace{-.3in} 
\caption[]{Average longitudinal hadronic shower profiles of neutrino events in
two different shower energy ranges.   
It can be seen from the plots that large fraction of shower energy is
	deposited in two to three consecutive counters.}  
\label{fig:long-profile} 
\end{figure}

The fitting procedure constrains the hadron energy of neutrino 
	interactions to be constant by varying the relative 
	gains of the counters.
Let the visible hadron energy of a neutrino event that starts 
	in counter $i$ be denoted as $EHAD_i$.    
In a given event the individual counter energies using the 
	muon-derived gains are denoted by $E(j)$, and the 
	hadron/muon gain ratio for each counter is denoted by $h_j$.  
In this notation, 
\begin{equation} 
EHAD_i = \Sigma_{j=i}^{i+9} h_j E(j)
\end{equation} 
where the sum over 10 counters (1~m of steel equivalent) 
is expected to include more than $95\%$ of the hadronic shower.  
	The average hadron energy over all neutrino events that 
occur in counter $i$ is then 
\begin{eqnarray*} 
<EHAD_i> &=& <\Sigma_{j=i}^{i+9}h_j E(j)>  \\
 &=&  \Sigma_{j=i}^{i+9}h_j <E(j)> 
\end{eqnarray*}
Thus in theory the average hadron energy of showers that start in 
	counter $i$ depends not only on counter $i$'s gain but also on
	the gains of the nine subsequent counters.  
In practice, however, hadron showers deposit a large fraction of 
	their energy in only two or three consecutive counters 
	around the shower maximum, as shown in 
	Figure~\ref{fig:long-profile}.  

One first computes the average hadron energy ($EAVE$) 
	over the entire fiducial volume by setting all the 
	initial gains $h_j$ to unity, making the cuts described 
	above, and calculating the average $EHAD_i$ over all 
	the events that pass the cuts.  
Then, one can fit for the hadron gains by minimizing a 
	$\chi^2$, where the $\chi^2$ is defined as follows:  
\begin{equation}\label{eq:hmu-gain-chi2}
\chi^2 = \Sigma_{i=20}^{80}\frac{(EHAD_i(g_i,g_{i-1},g_{i-2}...)-EAVE)^2 }
{ERR_i^2}
\end{equation} 
	and the $ERR_i^2$ is defined as: 
\begin{equation}
ERR^2_i = (<EHAD_i^2>-<EHAD_i>^2)/N_i  
\end{equation} 
	where $N_i$ is the number of neutrino events, passing all cuts, 
	that start at counter $i$, and the error is calculated assuming the 
	gains $h_j$ are all set to their initial values (which for 
	the first pass is simply unity).   
To fit for the gains, in theory, one simply has to minimize the 
	$\chi^2$ defined in Eq.~\ref{eq:hmu-gain-chi2}, determine the gains, 
	and then iterate, making the energy cut and re-computing the new 
	average and errors using the gains from the 
	previous iteration.  

In fact the $\chi^2$ defined in Eq.~\ref{eq:hmu-gain-chi2} is 
	unstable, due to the fact that a hadron deposits most of 
	its energy in two or three consecutive counters at the 
	shower maximum and this causes a strong correlation 
	between the two counters next to each other.  
The $\chi^2$, when computed this way, is low not only for 
	uniform gains very close to 1 but also for gains which 
	are staggered by a certain amount, where the even counters 
	are all high and the odd counters are all low (or vice versa).  
This variation in hadron gains is larger and more regular than 
	would be expected from detector non-uniformities in thickness 
	and composition of material. 
This artificial hadron gain variation is avoided by separately
	fitting the gains using 
events whose showers start in every other counter 
(for example, even-numbered counters), 
then using the complementary set of events (for example, 
showers starting in odd-numbered counters) and refitting.
The resulting gains for all counters are consistent between the 
	two fits, have smaller errors than when all showers are 
	included at once, and are much closer to unity.  

\begin{figure}[tbp] 
\begin{center}
\centerline{\psfig{figure=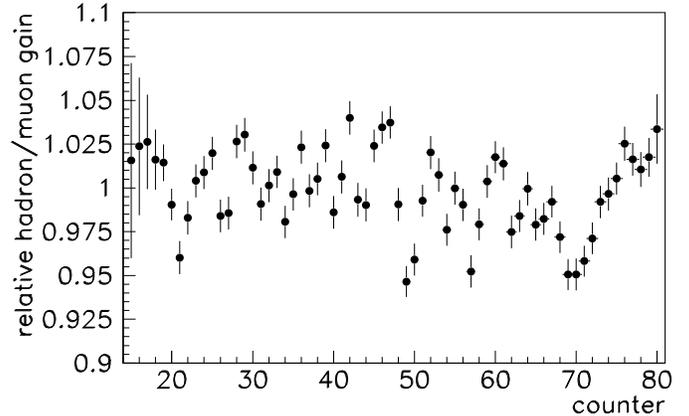,width=.5\textwidth}}
\end{center}
\vspace{-.3in} 
\caption[]{Relative hadron/muon counter gains which arise from detector
non-uniformities unrelated to scintillator thickness.}  
\label{fig:had_counter_gains} 
\end{figure}

By averaging the two fit results and iterating, the 
	gains are stable to better than $0.2\%$ after 3 iterations.  
The statistical uncertainty on each relative gain is about $0.9\%$, 
	and is larger near the downstream edge of the detector 
	where there are only events starting upstream of those counters.   
Figure \ref{fig:had_counter_gains} shows the gains obtained after 
	four iterations using the technique described in this section.  
The gains have an RMS of $2.3\%$, and are consistent with geometrical 
	non-uniformities in the calorimeter (water bag thicknesses, 
	steel plate thicknesses, etc.).  

The relative hadron gains allow a determination of the absolute 
	hadron energy scale of the entire calorimeter by measuring 
	the response of the upstream most counters to a monochromatic 
	beam of hadrons. 
The total statistical uncertainty on the hadron gains in the overlap 
	region where there is both calibration beam and neutrino data 
	is equal to $0.4\%$ and dominates the overall uncertainty in 
	the hadron energy scale.
Each counter's hadron/muon gain ratio uncertainty ($0.9\%$) is 
	uncorrelated between the counters.
The contributions of this uncertainty to the calibration beam 
	energy measurement is reflected in the uncertainty in 
	hadron response measurement and is negligably small
	due to statistically random longitudinal development 
	of hadron showers (see Section~\ref{sec:hadron}).

The relative gains obtained using the technique described 
	in this section are used for the energy reconstruction in 
	both the hadron and electron response measurements.

%
\section{The NuTeV Calibration Beam}
NuTeV is designed to include a simultaneous 
	calibration beam separated from the neutrino beam by 1.4 seconds, 
	yet running within the same one-minute accelerator cycle  
	(see Figure~\ref{fig:bm-time-structure}).
The calibration beam is used to set the absolute energy 
	scale of the experiment, and also to measure the response of the 
	calorimeter to hadrons, electrons, and muons, in order to 
	properly simulate the detector.
Finally, the calibration beam is instrumental in 
	monitoring the time dependence
	as measured by the muon map technique described in 
	Section~\ref{ss:gain_cor}.  

The calibration beam period within a cycle is 18 seconds, and the 
	typical beam incident angle to the center of the NuTeV calorimeter 
	is 43~mrad in the horizontal direction (0~mrad in vertical) 
	with respect to the centerline of the calorimeter.
The calibration beamline can transport particles of 
	energies from 4.8~GeV to 190~GeV, and depending on the beamline 
	apparatus and magnet settings, can produce high purity
	beams of electrons, hadrons, or muons 
for energies above 30~GeV.

The beamline is instrumented as a low mass spectrometer with a 
	long lever arm.
The distance between the most upstream chambers in 
	the spectrometer and the momentum-analyzing magnets is 83.3~m, 
	and the distance between the most downstream chamber and 
	the magnets is 69.2~m.  
This separation allows a modest alignment uncertainty of 1~mm to translate
	into only a 0.1\% uncertainty in the absolute momentum scale. 
The event-by-event resolution of the spectrometer, dominated by
	multiple scattering, is better than 0.3\% for most energies.
The beamline instrumentation is augmented for some of
	the run with a removable \u{C}erenkov detector and a TRD array used
	to measure the beam particle composition.

Over the course of the experiment, standard runs were taken at 
	least once a week (50 and 100~GeV hadrons) and hadron energy
	scans between 4.8~GeV and 190~GeV were taken once a month.  
Overall, NuTeV accumulated a total of 17 million test beam triggers.  

\subsection{Beam Time Structure}\label{ss:bm-time-structure}
\begin{figure}[tbp]
\begin{center}
\centerline{\psfig{figure=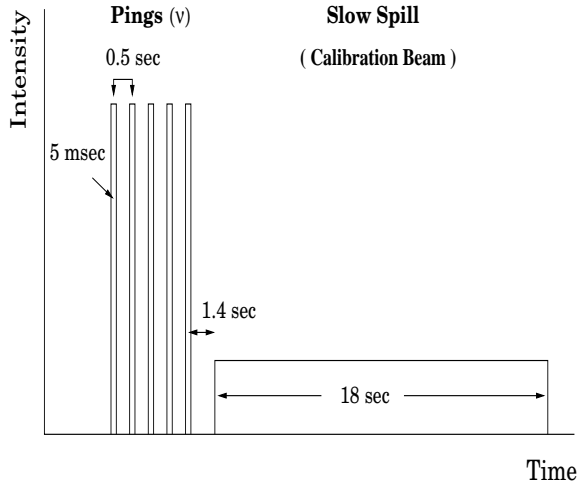,width=3.in,height=2.5in}}
\end{center}
\vspace{-.3in} 
\caption[]{Accelerator time structure.   Note that the interval 
	between the last neutrino ping and the slow spill calibration 
	beam is only 1.4 seconds, allowing essentially an 
	{\it in situ} calibration.}
\label{fig:bm-time-structure}
\end{figure}
The accelerator time structure during the 1996-97 Fermilab fixed
	target run is depicted in Figure~\ref{fig:bm-time-structure}.  
The accelerator complex cycles every 60.1~sec.  
The neutrino beam is delivered in five fast resonance extraction 
	pulses (``pings'') of 5~msec width.
The pings are separated by 0.5~sec.  
The NuTeV slow spill calibration beam begins 1.4~sec after the last 
	ping and has a duration of 18~sec with uniformly
	distributed beam intensity.
This calibration beam is delivered in a beamline that is completely
	independent from the fast spill.  
This time structure provides continuous calibration data, taken 
	concurrently with the neutrino beam, and allows for an 
	{\it in situ} calibration of the detector.
\subsection{Beam Selection Scheme}\label{sec:beamselect} 
\begin{figure}[tbp]
\begin{center}
\centerline{\psfig{figure=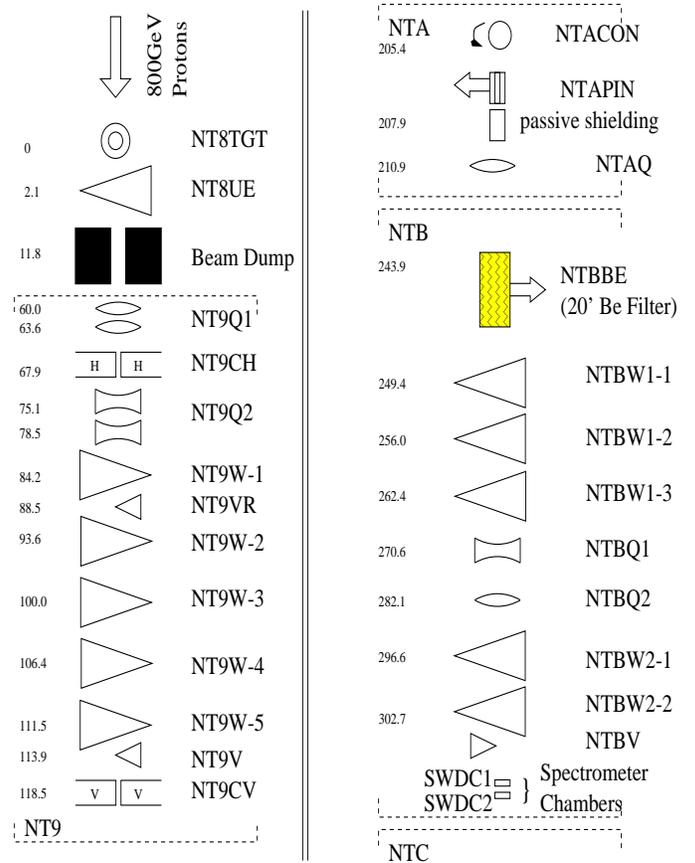,width=3.5in,height=4.5in}}
\end{center}
\vspace{-.3in} 
\caption[]{NuTeV calibration beamline schematics.
The ``Ferris wheel'' (NTACON) with four 
	different thickness converter material is used to select 
	pure hadrons or electrons.
The 6~m long Be filter(NTBBE) is used to select pure muons.
The numbers on the left-hand-side of each component indicate 
	the relative distance of the component to the primary 
	target (NT8TGT) in meters.
Some beam position and intensity monitoring devices are not drawn in
	this figure because they are irrelevant for this paper.
}
\label{fig:nt-bm-line}
\end{figure}

The NuTeV calibration program involves electrons, hadrons, and muons of
	momentum ranging from 4.8~GeV to 190~GeV.
In order to select the desired
type of particles for a specific program with
	high purity, the beamline is designed as shown in 
	Figure~\ref{fig:nt-bm-line}.

The target (NT8TGT) in the calibration beam is a
	$7.5~{\rm cm~(W)}\times 7.5~{\rm cm~(H)}\times 30.3~{\rm cm~(L)}$ 
	aluminum block.
Protons of 800~GeV momentum strike the target with an integrated 
	intensity between
	$4\times 10^{11}$ and $8\times 10^{11}$ protons on target 
	throughout the 18 second long slow spill.
The secondaries are then focused by a set of quadrupole 
	magnets (NT9Q1 and NT9Q2) to the enclosure NTA and collimated 
	by a horizontal collimator (NT9CH) whose
	opening is adjusted depending on particle type and intensity.
The polarity of the beamline is set to direct negatively charged 
	particles to reduce intensity.

The horizontal collimator (NT9CH) is then followed by a string of 
	dipole magnets (NT9W-1, 2, 3, 4, and 5), fed by one power supply,
 	for initial momentum selection, bending the beam in the 
	horizontal plane.
The vertical collimator (NT9CV) following the 
	first set of dipoles (NT9W's) is used to further
	cut down the intensity.

The ``Ferris wheel'' (NTACON), located immediately downstream of the 
	set of collimators and the initial momentum selection dipoles, 
	has four mounts, each placing a different thickness
	of material into the beam at a time.
The thicknesses correspond to an empty hole, 0.2$X_{0}$, 6$X_{0}$, 
	and 12$X_{0}$.
The empty hole is used for the muon mode, while the 0.2$X_{0}$ 
	piece is used for the electron mode.
In the electron mode, the magnets downstream of the ``Ferris wheel''
	are tuned for 20\% lower momentum particles to compensate
	electron energy loss in the material.
The two higher radiation length materials are put in the path of the beam
	for hadron modes to eliminates electrons from the beam.  The 
thicker material is used for higher energy beams, the thinner for lower.   

The pinhole collimator (NTAPIN) following the ``Ferris wheel'' is put in the 
	beam only for higher energy hadron modes ($E>30$~GeV) to further cut 
	down intensity and increase radiation safety.
Typically, the size of the hole in the pinhole collimator is 
	$5~{\rm mm}\times 5~{\rm mm}$ and the momentum bite set by 
	this collimator opening is approximately 0.2~GeV.

The 6~m long beryllium filter (NTBBE) is only used in muon modes to filter 
	out hadrons and electrons in the beam. 
The energy loss of muons in the filter is approximately 6\% while the 
	survival probabilities of hadrons and electrons through the 
	filter are less than $3.4\times 10^{-7}$ and 
	$3.5\times 10^{-8}$, respectively.
The filter is then followed by additional two sets of dipoles 
	(NTBW1-1, 2, 3, NTBW2-1, and 2) for further 
	refinement of momentum selection.
\begin{figure}[tbp]
\begin{center}
\centerline{\psfig{figure=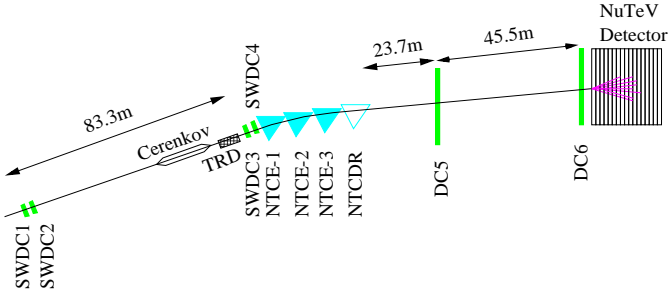,width=3.5in,height=1.5in}}
\end{center}
\vspace{-.3in} 
\caption[]{A schematic view of the NuTeV calibration
	beam spectrometer system.   
The large distances between the chamber 
	stations allow accurate absolute momentum determination.}
\label{fig:spectrometer}
\end{figure}
This combination of three large dipole strings throughout the 
	long stretch of the beamline removes virtually all possible 
	contamination of unwanted particles and momenta.

A final precision spectrometer is used to determine the beam momentum
	on an event-by-event basis.
The spectrometer begins with two small area drift chambers 
	(SDWC1 and SDWC2 in Figure~\ref{fig:nt-bm-line}) 
	positioned at the downstream end of the last dipole 
	string in the same beam enclosure (NTB). 
Figure~\ref{fig:spectrometer} shows a schematic view of the 
	NuTeV calibration beam spectrometer system.
The particle ID system, which consists of a \u{C}erenkov counter followed
	by an array of TRD's, is located just upstream of the second set 
	of chambers which,
in turn, were positioned immediately upstream of the spectrometer 
	dipole magnet string.
When particle identification is not needed, these detectors are 
	rolled out of the beamline and are replaced by a vacuum 
	pipe to reduce multiple scattering.

The last dipole in the spectrometer magnet string can be rotated.  
This dipole is an integral part of the spectrometer for beams with 
	energies greater than or equal to 120~GeV and is also used to 
	direct the beam to various positions on the detector surface for 
	position dependent response measurements.

\begin{figure}[tbp]
\begin{center}
\centerline{\psfig{figure=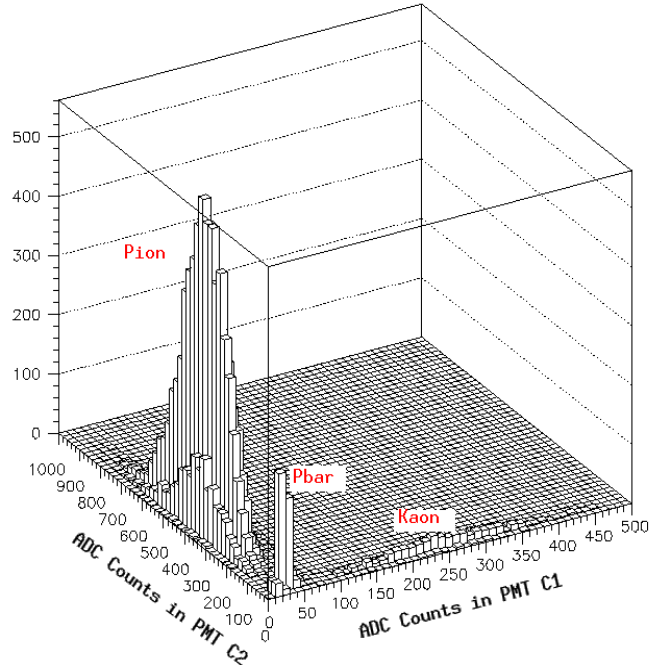,width=3.5in,height=3.5in}}
\end{center}
\vskip -0.2in
\caption[]{
PMT signals of the \u{C}erenkov counter with a 160~torr 
	nitrogen gas.  
A clean particle separation between $\pi$($C_1$), K($C_2$), 
	and $\pbar$(ped) at 50~GeV is apparent.}
\label{fig:50gev_demo}
\end{figure}
\subsection{Particle Identification and Beam Purity}
The \u{C}erenkov counter provides particle identification 
	for pions, kaons, anti-protons, and electrons, 
	depending on the type of gas and
	the threshold pressure for each type of particles. 
The \u{C}erenkov counter is equipped with two PMT's, 
	$C_1$ and $C_2$, placed to face opposite directions, 
	and is designed to act as a differential 
	\u{C}erenkov counter.
The PMT $C_1$, accepts low angle ($<4.5$~mrad) 
	\u{C}erenkov light from heavy particles, while the 
	PMT $C_2$, accepts large angle light from 
	lighter particles of the same momentum.

Figure~\ref{fig:50gev_demo} demonstrates the excellent particle 
	identification for anti-protons, kaons, and pions within the
	50~GeV hadron beam using the \u{C}erenkov counter under 
	nitrogen, at a pressure of 160~torr.  
While the small signal in the pedestal region is dominated by 
	anti-protons, it could also be contaminated by other
	particles due to inefficiencies in $C_1$ and $C_2$. 
An inefficiency study, performed by counting the number
	of pedestal events for the clean muon sample with 
	\u{C}erenkov pressure above the muon threshold, shows that 
	the $C_1$ and $C_2$ inefficiencies are less than 
	0.24\% and 0.008\%, respectively.

\begin{table}[tbp]
\begin{center}
\begin{tabular}{|c|l|l|}\hline\hline
 P     &\hfil Electron &\hfil Hadron fraction(\%) \\ 
 (GeV)   &  fraction(\%) & 		  \\ \hline\hline 
    5    &      92	&     8		 \\
  7.5	 &	72	&	28	 \\
  15    &	66	& 34 ($\pi^{-}$:95.6, $\pbar$+${\rm K^{-}}$:4.1)\\
  20	&  $<$ 1	& $>$ 99 ($\pi^{-}$:95.5, $\pbar$+${\rm K^{-}}$:4.5) \\ \hline
  30	&  $<$ 0.25	& $>$ 99.75 ($\pi^{-}$:94.9,$\pbar$+${\rm K^{-}}$:5.1) \\ \hline
  50	&	0	& 100 ($\pi^{-}$:93.9, ${\rm K^{-}}$:3.1, $\pbar$:3.0)\\ \hline
  75	&	0	& 100 ($\pi^{-}$:91.7, ${\rm K^{-}}$:5.1, $\pbar$:3.2)\\ \hline 
 120	&	0	& 100 ($\pi^{-}$:91, ${\rm K^{-}}$:6.2, $\pbar$:2.8)\\ \hline\hline
\end{tabular}
\caption[]{Summary of particle composition ($e^{-}$/$\pi^{-}$/${\rm K^{-}}$/$\pbar$) in 
	the hadron calibration beam for various energies.}
\label{tb:tb-fraction}
\end{center}
\end{table}
Table~\ref{tb:tb-fraction} summarizes the beam particle composition for
	various hadron tunes.
It is well demonstrated that the contamination from electrons in the
	beam for hadron tunes of momentum greater than 30~GeV is less
	than 0.25\%, minimizing the systematic error in hadron response
	measurements.
\subsection{The Calibration Spectrometer}
The spectrometer is designed to measure  the absolute momentum of
the calibration beam particles to better than 0.3\%
	in an event-by-event basis. 
This is accomplished by two means.  First 
precisely calibrated dipole magnets are used, with 
	$\int Bd\ell$ known to better than 0.1\% in 
	the region traversed by the beam. Also, 
the bend angle is measured to 
	better than 0.1\% using drift chambers positioned over the 
	500~m beamline, which provides a long lever arm.
This long length of the spectrometer chamber spacing allows us 
	to tolerate a relative chamber alignment uncertainty 
of $\sim$ 100~$\mu{\rm m}$. 
\subsubsection{Upstream Tracking}
\begin{figure}[tbp]
\begin{center}
\psfig{figure=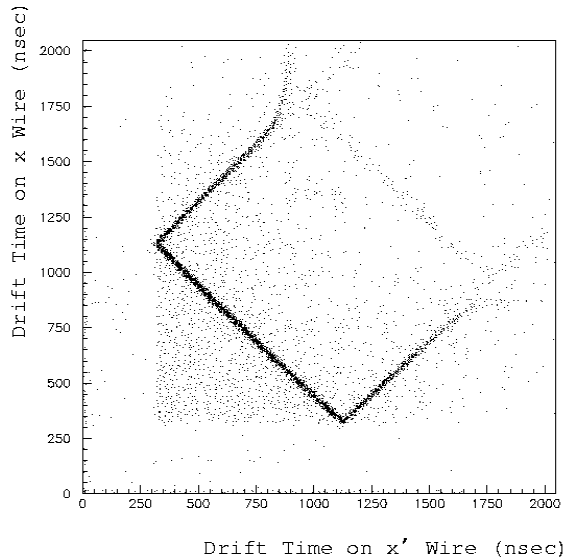,width=3.in,clip=t,height=3.in}
\end{center}
\vspace{-.3in} 
\caption[]{$T_1$ vs $T_2$ for $x-$view of an SWDC  plane. The cluster
of points along the line with slope $-1$ to the left corresponds to the
region between the $x$ and $x^{\prime}$ wires; these events are used for
track fitting. Nonlinear time-to-distance effects are visible for very long
drift times and for events very close to the sense wires. }
\label{Krider T1 vs T2}
\end{figure}

The position and the angle of the calibration beam tracks are 
	determined at the upstream
	end of the spectrometer magnet by four 
	$12~{\rm cm}\times 12~{\rm cm}$ Single Wire Drift 
	Chambers (SWDC)~\cite{krider}.
Each chamber consists of a pair of sense wires 
	offset by $\pm 2.03$~cm from the beam center in each view. 
The operating gas, an equal mixture of Ar and C$_2$H$_6$, and 
	the field-shaping wires provide a saturated $49~\mu{\rm m/ns}$ 
	drift speed over most of the gas volume. 
This can be seen in Figure~\ref{Krider T1 vs T2}, where the drift
	times measured on the two sense wires are plotted against 
	each other. 
The dark band with slope $-1$ is produced by tracks passing between the two
	wires. 
The two bands with slope $+1$ are produced by the tracks passing on the
	same side of the two wires. 
Small non-linear drift effects can be seen at very long and very 
	short drift times. 
These effects are eliminated by using only tracks passing between 
	the two wires.

The chambers are grouped in two stations of two SWDC's each.
One station is located immediately upstream of the most upstream 
	spectrometer magnet and the other is 83.3~m upstream of 
	that station. 

\begin{figure}[tbp]
\begin{center}
\epsfxsize=0.45\textwidth
\epsfbox[81 240 500 666]{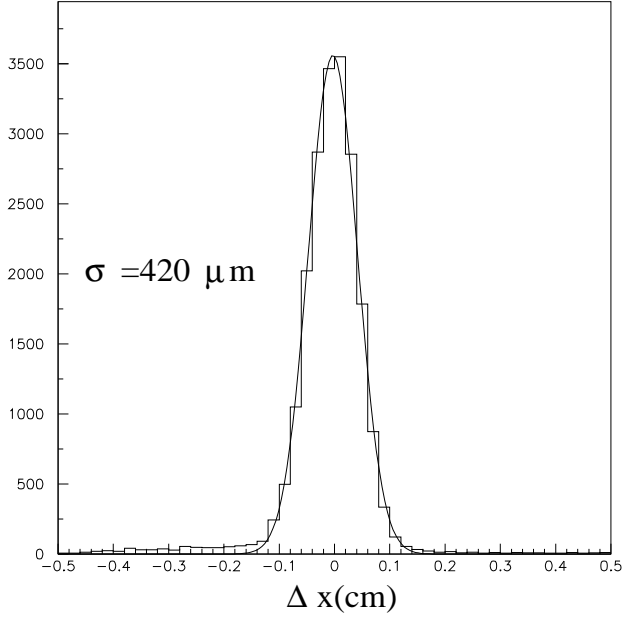}  
\end{center}
\caption[]{ Difference in coordinates, $X_1-X_2$ for an SWDC chamber plane
from a sample of calibration beam tracks. 
The solid line represents a Gaussian fit on the distribution.
The $\sigma $ of this distribution is 420~$\mu{\rm m}$, implying a spatial 
	resolution of 300~$\mu{\rm m}$. 
}
\label{Krider resolution}
\end{figure}

Chamber position resolution can be determined by measuring  the width of the
	distribution of differences in position measurements for a track 
	passing between the two sense wires, and then dividing by $\sqrt{2}$. 
This is shown in Figure~\ref{Krider resolution}, which demonstrates that the
	chambers have a resolution of $300~\mu{\rm m}$. 
Chamber alignment is achieved through an initial optical survey to 
an accuracy of $50~\mu{\rm m}$. 
The integrity of the alignment is constantly checked using the 
	straight test beam tracks, with the spectrometer magnets
 	removed from their normal positions.
Residual misalignments are estimated to be $\leq 100~\mu{\rm m}$ and
	 make a negligible contribution to 
	slope and intercept measurements.
\begin{figure}[tbph]
\begin{center}
\epsfxsize=0.45\textwidth
\epsfbox[90 240 500 666]{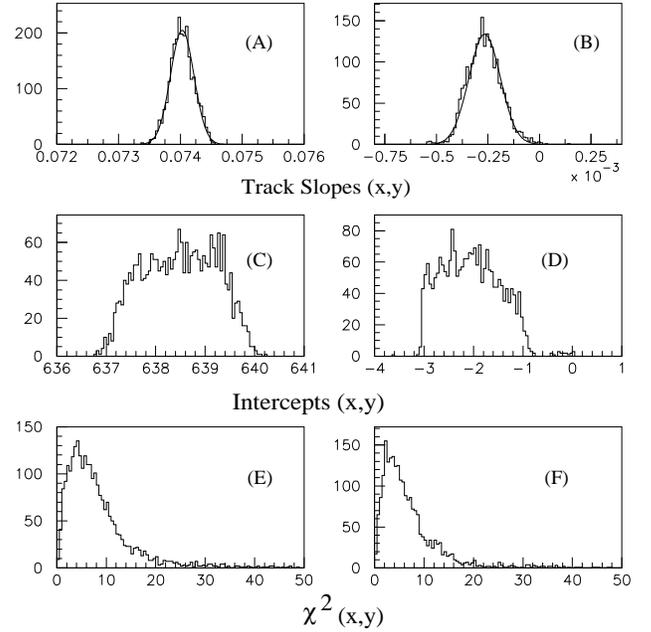}  
\end{center}
\caption[]{ 
A and B: Track slopes in $x$ and $y$ at the upstream end of calibration
	 spectrometer magnets as determined by chamber tracking.
C and D: Beam profile in $x$ and $y$ at the upstream end of calibration
	 spectrometer magnets as determined by chamber tracking.
E and F: $\chi^2$ distribution of the upstream chamber fits in 
	 $x$ and $y$ views.  Both distributions
	 follow the expected normal $\chi^{2}$ distributions.
}
\label{upfits}
\end{figure}
\begin{figure}[tbp]
\begin{center}
\epsfxsize=0.45\textwidth
\epsfbox[65 240 500 666]{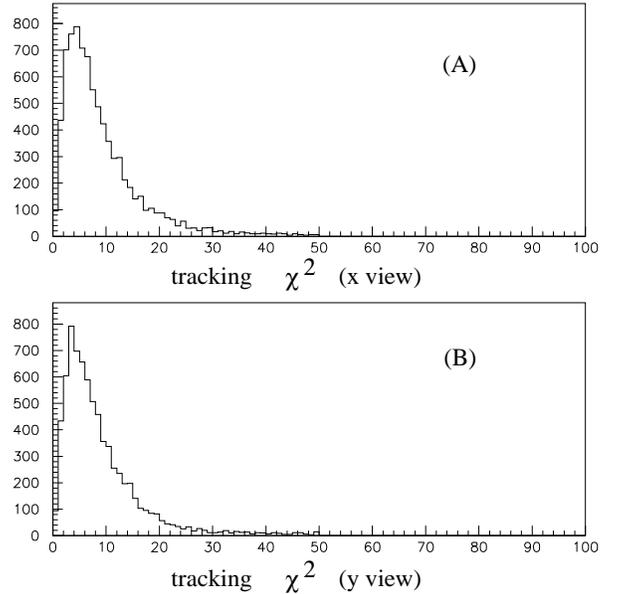}  
\end{center}
\caption[]{
$\chi^2$ distribution for X (A) and Y (B) view of 
	calibration beam spectrometer track fit using 
	chambers upstream and downstream of 
	the spectrometer magnet string. 
Ten points are used in the three parameter fit. }
\label{UD chisq}
\end{figure}

Figure~\ref{upfits} shows the distribution of slope, intercept, and $\chi ^2$
	for a typical upstream sample of tracks in each view. 
Tracks enter the spectrometer at projected angles of $\theta _x=74$~mrad, 
	$\theta _y=-0.3$~mrad with angular spreads of $\sigma _x=0.2$~mrad, 
	$\sigma _y=0.08$~mrad. 
The width of the beam is approximately 2.5~cm in $x$ and $y$ and is set by
	the trigger paddles. 
The $\chi^2$ distributions are consistent with their expected shape, 
	giving us confidence that the upstream tracking is well understood.
\subsubsection{Downstream Tracking}
The downstream section of the spectrometer consists of two 
	$3~{\rm m}\times3~{\rm m}$ three-wire drift chambers 
	separated by 45.5~m.
The first chamber downstream of the spectrometer magnet 
	string is positioned 23.7~m from the downstream end of the 
	last momentum analyzing magnet.
Figure~\ref{UD chisq} shows the $\chi^{2}$ distributions of the
spectrometer tracking fit, using the chambers upstream
	and downstream of the spectrometer dipole magnets.
The fact that these distributions follow the expected $\chi^{2}$ distribution
	gives us confidence in the absolute momentum determination.
\subsubsection{Spectrometer Magnet Calibration}
The Fermilab Magnet Test Facility (MTF) calibrated the five EPB 
	dipoles~\cite{EPB dipole} (four plus an unused spare) 
	used in the spectrometer.
\begin{figure}[tbp]
\begin{center}
\psfig{file=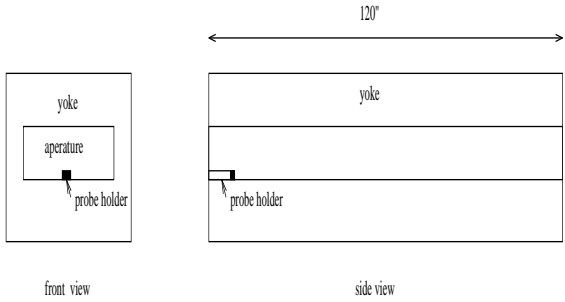,clip=,height=2.in,width=3.in}
\end{center}
\vskip -0.6in
\caption[]{ 
A schematic diagram of the Hall probe and the holder 
	location inside the magnet.
}
\label{holder}
\end{figure}
Precise $\int Bd\ell$ data are taken at the centerline of the magnet 
	and are tied to magnet current and Hall probe voltage 
	readout recordings.
Shape studies are performed for $\int Bd\ell $ vs horizontal 
	position at fixed vertical position and magnet shunt
	current measurements~\cite{EPB procedure}. 
Data are summarized by polynomial fits to the $\int Bd\ell$ measurements 
	as functions of both Hall probe output and magnet current. 

While shunt devices can be internally calibrated to better 
	than 1 part in 10$^4$, the current reading in
	the two different configurations (i.e. different power 
	supplies, buses, cables, and shunts) may differ by 
	substantially more
	and cannot be used to obtain the absolute $\int Bd\ell$. 
Therefore, the absolute $\int Bd\ell$ is determined in the data taking 
	configuration based on the Hall probe versus $\int Bd\ell$ 
	calibration data. 
Further details of the magnet calibration are described 
	elsewhere~\cite{magnet memo}.
\begin{table}[tbp]
\begin{center}
{\small
\begin{tabular}{|l|l|l|l|l|}
\hline
\textbf{Magnet} & \textbf{Probe} & $\mathbf{A}_1$ (m) & $\mathbf{A}_0$ (T$%
\cdot $m) & $\mathbf{A}_2$ (T$^{-1}\cdot $m) \\ \hline
11243 & 95421 & -3.03617 & -0.00517 & 0.00574 \\ 
11243 & 95420 & -3.04009 & -0.00438 & 0.00283 \\ 
11243 & 95422 & -3.03576 & -0.00448 & 0.00130 \\ \hline
11243 & 95423 & -3.03392 & -0.00276 & 0.00056 \\ \hline
11459 & 95421 & -3.03770 & -0.00491 & 0.00046 \\ \hline
11632 & 95421 & -3.03575 & -0.00532 & 0.00078 \\ \hline
11694 & 95420 & -3.03983 & -0.00351 & 0.00066 \\ \hline
20015 & 95421 & -3.04207 & -0.00473 & 0.00051 \\ \hline
\end{tabular}
}
\end{center}
\caption[]{
Coefficients of fits to $\int{Bd\ell}$ vs Hall probe readout for 
different calibration
spectrometer EPB dipole/Hall probe combinations. The fits are of the
form $\int{Bd\ell}=A_0+A_1H+A_2H^2$, where $H$ is the Hall probe readout in
Tesla and $\int{Bd\ell}$ is in units of Tesla-meters.}
\label{Hall fits}
\end{table}

Each of the four Hall probes is attached to a probe holder before the
	holders are mounted in the magnets. 
The probe holders are located approximately in the center 
	magnet aperture on the lower pole face near the magnet opening. 
This is shown schematically in Figure~\ref{holder}. 
Holders are angled to keep the probe cables from interfering with 
	the beam and vice versa. 
Hall probes are read out and the values are recorded by the NuTeV data 
	acquisition system once per spill, and $\int Bd\ell $ 
	data are calculated for all events within a given spill 
	using quadratic fits to the MTF Hall probe calibration data. 

\begin{figure}[tbp]
\begin{center}
\psfig{file=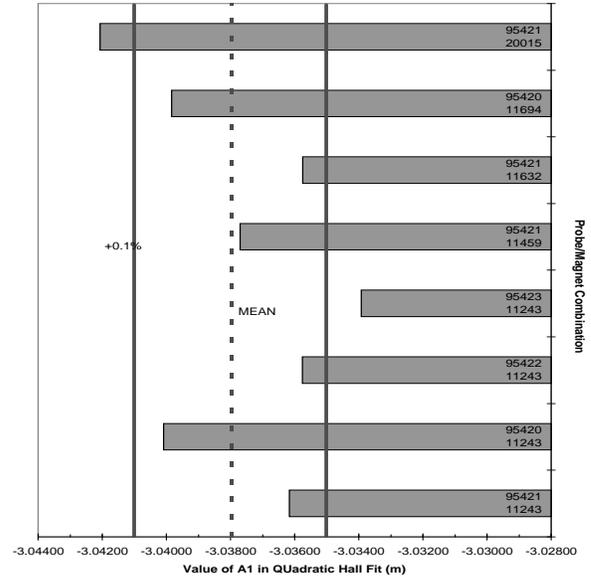,clip=,height=3.in,width=3.in}
\end{center}
\vspace{-.3in} 
\caption[]{ 
Comparison of linear coefficients in 
$\int{Bd\ell}$ vs Hall probe readout for different 
EPB dipole/ Hall probe combinations for the NuTeV
calibration beam spectrometer magnets.
}
\label{Hall calibration}
\end{figure}
\begin{figure}[tbp]
\begin{center}
\psfig{file=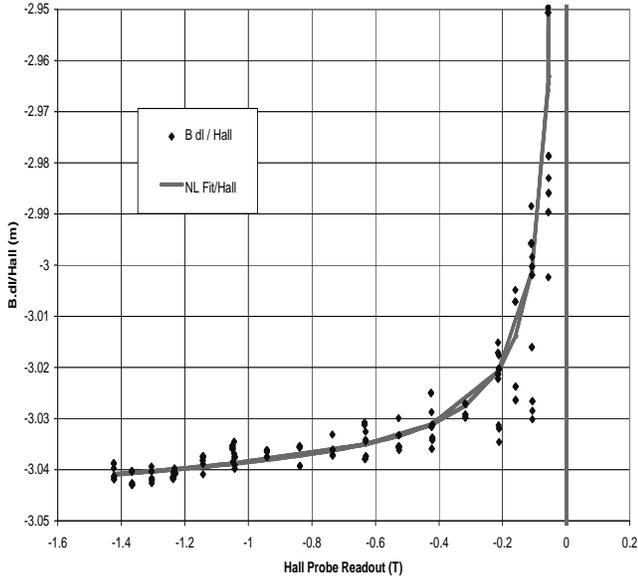,clip=,height=3.in,angle=270,width=3.3in}
\end{center}
\caption[]{ 
Results of quadratic fit to $\int{Bd\ell}$ vs Hall probe readout for a 
typical EPB dipole in the NuTeV calibration spectrometer.
$\int{Bd\ell}/H$ vs $H$, where $H$ is the Hall probe reading in 
Tesla and $\int{Bd\ell}$ in Tesla-meters is plotted to show 
the non-linear region at low fields.
}
\label{Hall resids}
\end{figure}

Table~\ref{Hall fits} summarizes the $\int Bd\ell $ vs Hall probe fits to
	the data. 
The $\int Bd\ell $-Hall probe relationship is very nearly linear,
	with the offset and quadratic corrections 
	($A_0$ and $A_2$ in Table~\ref{Hall fits}) only a few 
	parts-per-mil of the linear calibration constants $A_1$. 
Figure~\ref{Hall calibration} compares the coefficient $A_1$
	for different probe-magnet combinations; variations from 
	dipole-to-dipole and probe-to-probe are at the few tenths 
	of a percent level.
Linear fits do not parameterize the data to the required accuracy 
	($\leq 0.1\%$ deviation), but quadratic fits 
	well describe both polarities of current ramping. 
Figure~\ref{Hall resids} shows the fit results superimposed on the 
	data for one of the magnets; plotted are the $\int Bd\ell $ 
	points divided by the probe readout in order to accentuate 
	non-linear effects.

\begin{figure}[tpbh]
\begin{center}
\epsfxsize=0.45\textwidth
\epsfbox[90 240 500 660]{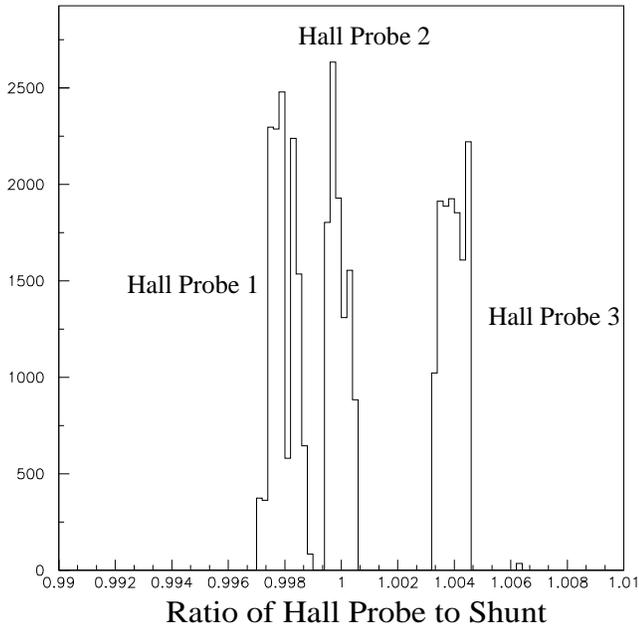}
\end{center}
\caption[]{ 
Ratio of $\int Bd\ell$ calculated from magnet shunt current to the value
	calculated from the Hall probe for a typical NuTeV calibration run.
}
\label{HP vs I}
\end{figure}
As a check, $\int Bd\ell $ values have also been calculated from high-order
	polynomial fits to $\int Bd\ell $ vs magnet shunt current 
	data taken at MTF.
Figure~\ref{HP vs I} compares the Hall probe determination to the shunt
	current determination of three spectrometer magnets 
	for a typical run.  
The two determinations agree within the expected precision of 
	the shunt current measurement.
\subsection{The Calibration Trigger}
\begin{figure}[t]
\begin{center}
  \epsfxsize=.3\textwidth
  \epsfbox{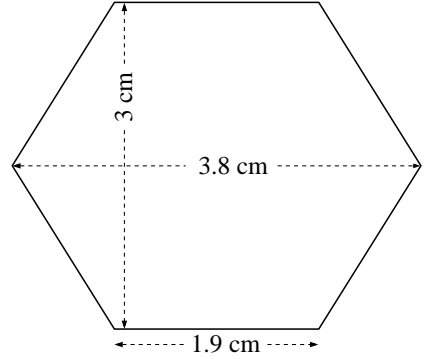}
\end{center}
\caption[]{A schematic drawing of the NuTeV calibration beam 
	trigger scintillation paddle.}
\label{fig:trigger-paddle}
\end{figure}

The calibration beam trigger consists of two small scintillator paddles
	shaped to shadow the ``good field'' regions of the spectrometer
	magnets.
The two paddles are positioned immediately upstream and downstream 
	of the momentum analyzing magnet string. 
Figure~\ref{fig:trigger-paddle} shows a schematic diagram of a
	calibration beam trigger scintillator paddle.
The ``good field'' region, mapped out with $\int Bd\ell$ 
	measurements described above, consists of the region across the 
	face of the magnet over which the $\int Bd\ell$ varies by less than
	0.1\% from its value at the center of the magnet. 
This unbiased trigger, with no energy requirement in the calorimeter,
	automatically maintains the 0.1\% tolerance on the $\int Bd\ell$.
%

%
\section{Corrections for Systematic Effects}
In order to achieve sub-per cent precision in the absolute
	energy scale calibration, it is necessary for NuTeV to take
	into account a number of systematic effects.  
Most of these are due to the fact that the gains of the calorimeter 
	are determined using muons from the neutrino beam which
	averages the detector response over a long period of time
	(typically a week or more), while the calibration beam runs
	take place over much shorter time periods; for example, some
	hadron energy tests run for as little as an hour.  
The time dependent effects that need to be taken into account and 
	that are discussed in this section are 
	high voltage and temperature gain dependencies.  

There are other differences between the energy 
	deposition from a neutrino interaction and 
	that from an incoming beam of particles.  
One of the differences is the calibration beam composition.
The particle type dependence of the energy deposition is studied, and
	correction for the anti-proton contamination is applied for 
	final energy scale calibration.

Another difference between neutrino and calibration beam energy deposition
	arises from the fact that a neutrino may interact 
	at different distances from a scintillation counter, whereas the
	calibration beam always enters the calorimeter at the front and, 
	in particular, electrons always interact in the first counter, 
	which is preceded by two inches of steel.  
For this reason special care
	is taken to ensure that the calibration of the first few
	counters in the calorimeter is consistent with that of the latter
	ones.
%
\subsection{Environmental and Voltage Monitoring System}\label{ss:slow-mon}
In order to obtain corrections for systematic effects, 
	temperature, pressure, humidity, high voltages, and low voltages
	are monitored locally by a microprocessor controlled system.
	This system communicated periodically with the data acquisition 
	computer to record the monitored data on the same
	tape as the neutrino and calibration data.  
The period for recording this data is one beam cycle, 
	about one minute.

The microprocessor signals were digitized by a 12-bit ADC
 	and then read into a Basic Stamp BS2 microprocessor. 
The signal is averaged over many readings to avoid noise. 
The results of this averaging were transmitted via a standard serial 
	communication network to a personal computer in the control
	room. 
The personal computer monitored these values, issued warnings when
	values began to deviate from the standard, issued alarms 
	when the values were out of limits, and transmitted the raw 
	data to the data acquisition system to be written to tape. 

The temperature of the calorimeter is monitored with a
	digital temperature integrated circuit to a resolution 
	of $\pm 0.6^\circ C$.  
The temperature probes are placed in four locations on every 
	fourteenth calorimeter counter unit. 
The absolute atmospheric pressure is monitored in four locations
	with a resolution of $\pm 1.5\%$. 
The other environmental variables are monitored to a 
	resolution of \hbox {$\pm 2\%$} in several places 
	throughout the experimental hall. 

The calorimeter PMT high voltages are monitored using the LeCroy 1440
	main frame readout system.
The readout is recorded once per cycle to the data tape together 
	with neutrino and calibration data.

\subsection{Beam Component Correction}
The absolute hadron energy scale of the calorimeter
is determined by measuring its response to single pion interactions.  
Any difference in the response of the detector between pions and 
the kaons or anti-protons which contaminate 
the calibration beam must be accounted for 
when the absolute energy scale is set.
These differences in the calorimeter response are investigated
	using clean samples of each particle type using the
	\u{C}erenkov counter information.

\begin{table}[tbp]
\begin{center}
{\small
\begin{tabular}{|c|l|l|l|}\hline\hline
    P   &\hfil shower response      &\hfil fraction  &\hfil correction \\ 
 (GeV)  &     (normalized to pion)  &   (\%)         & to $E_{had}$ (\%) \\ \hline\hline
   5    & $\pbar$ 1.2 	            & 3$\pm$1	 & -0.6$\pm$0.2  \\ \hline 
  10    & $\pbar$ 1.1               & 3$\pm$1	 & -0.3$\pm$0.1  \\ \hline
  15    & $\pbar$+${\rm K}^{-}$: $1.054\pm0.017$& 4.1	 & -0.22 \\ \hline
  20	& $\pbar$+${\rm K}^{-}$: $1.033\pm0.010$& 4.5	 & -0.15 \\ \hline	
  30	& $\pbar$+${\rm K}^{-}$: $1.027\pm0.006$& 5.1	 & -0.14 \\ \hline	
  50	&  $\pbar$: $1.011\pm0.006$ & 3.0	 & -0.017 \\		
	&    ${\rm K}^{-}$: $0.995\pm0.006$	    & 3.1	 & \\ \hline
  75	& $\pbar$: $1.008\pm0.004$  & 3.2	 & -0.010 \\		
	&    ${\rm K}^{-}$: $0.997\pm0.004$     & 5.1	 & \\ \hline
 120	& $\pbar$: $1.005\pm0.004$  & 2.8	 & -0.002 \\
	&    ${\rm K}^{-}$: $0.998\pm0.003$     & 6.2	 &  \\ \hline\hline
\end{tabular}
}
\caption[]{Hadronic shower responses from kaon and anti-proton normalized to that of
	pions, and the correction factors to the hadronic shower energy, 
        especially due to the anti-proton effect.}
\label{tb-response}
\end{center}
\end{table}
Based on the studies using the hadron beam at various energies, we 
	find that the calorimeter response to kaons agrees with that 
	of pions.
However, showers from anti-protons show higher responses
	(by $\sim$1~GeV) than the showers from pions,
	due to the $p \pbar$ annihilation at the end of 
	shower development process.
This effect has been discussed in a previous calorimeter 
	review~\cite{ex:anti-proton}.

At high energies ($\geq$ 50~GeV), the anti-proton effect is 
	found to be negligible ($<$ 0.03\%), while this effect 
	is very important at low energies.
Table~\ref{tb-response} summarizes the sizes of 
	the correction factors, 
	the contamination of kaons, anti-protons, and the shower 
	responses (normalized to the pion shower response) of
	the calorimeter.
\subsection{Muon Radiative Equilibrium (RE) Correction}\label{sec:eqcor} 
When a muon traverses material, it loses energy 
	via electromagnetic processes: knock out electrons 
	($\delta$-ray) from atoms, bremsstrahlung, 
	$e^{+}e^{-}$ pair production, etc.
While most of the knock-on electrons are low energy electrons
	that do not penetrate deep into the material, 
	high energy electrons from muon energy loss processes 
	can leave energy in several counters. 
Thus, the energy deposited in the most upstream few counters in 
	the NuTeV calorimeter is relatively lower than other 
	downstream counters since they have less material 
	in before them.
This effect is called the radiative equilibrium (RE) effect.
Since the NuTeV calibration beam enters the detector striking the most 
	upstream counters, and the gain corrections for the 
	counters are determined relative to muon energy deposit 
	in a given counter, it is necessary for analyses to 
	apply corrections to the gain factors to account 
	for the RE effect.
This effect causes an artificial over-estimate of the energy 
	deposited in a few upstream counters relative to the 
	downstream ones.

We determine the size of this correction using a high 
	statistics GEANT Monte Carlo study.
Since this effect reduces the muon energy deposit 
	gain normalization factors for the most upstream and 
	the second most upstream counters by 1\% and 0.4\%, 
	respectively, the normalized hadron energy deposit 
	in these two counters need to be reduced by the
	same factors. 
The resulting overall size of this correction to the
	hadronic response is less than 0.1\%.
\subsection{Temperature Correction}\label{ss:temp-cor}
Many characteristics of the NuTeV calorimeter -- PMT high voltage, PMT 
	quantum efficiencies, scintillator light yield, electronics 
	noise, etc -- change with temperature.
These changes contribute to the temperature dependence of the  
	overall gain.
Separating systematic effects from these different sources 
	is difficult and unnecessary.  
The net effect that the temperature has on the
	overall gains of the PMT's can be measured 
	by muon response maps and the average 
	temperatures for each of these muon maps.
That temperature dependence is shown in Figure~\ref{fig:gain_vs_temp} in
	Section~\ref{ss:gain_cor}.

The muon map of the counter for a given running period
	is the time-integrated, beam-weighted response of the 
	counter for a particular period of time.  
Since the neutrino data itself is, by definition, beam-weighted, the 
	``average temperature'' for a muon map is also 
	the average temperature for the neutrino data.  

However, this is not true for calibration beam data.
A particular calibration beam study might only take a few
	hours, while the average muon map is calculated over a few weeks.  
If, during the few hours of the calibration beam study, the temperature
	is significantly different from the average temperature for 
	the muon map used, the gain of the calorimeter during 
	that brief time interval would be 
	different from the muon map average gain.  
To correct for these gain differences, a temperature correction 
	is applied to the calibration beam data on an event-by-event 
	basis, such that the effective muon map used would be 
	the appropriate muon map for that particular temperature.  

To calculate the corrections, the average gain for each counter 
	($G=\frac{g_A+g_B+g_C+g_D}{4}$)
	is linearly correlated to the measured temperature 
	from the sensors located on the calorimeter 
	($G(T)=A\times T+B$).  
Although there are sensors placed along the length
	of the calorimeter and the temperatures are 
	measured throughout the experimental hall, 
	the temperatures measured by the sensors near 
	the least insulated part of the calorimeter 
	are used to determine the temperature correction.
The temperature correction to each counter is simply the 
	first term in the Taylor expansion of ratio 
	of $ G(T_{muon~map})/G(T_{current}) $, or 
	($1-\frac{B}{A}\times(T_{muon~map} - T_{current})$).  
Temperature corrections tend to be as large as several tenths 
	of a percent.

\subsection{High Voltage Correction}
Counters in the NuTeV calorimeter have four PMT's, one in 
	each corner, as described in Section~\ref{ss:detector}.
The overall gain of a given counter depends strongly on the 
	combination of the individual PMT gain.
One of the systematic factors that directly affects the gain is
	PMT high voltage (HV).
Thus, for a high precision calibration, it is important to correct
	for overall gain fluctuations due to any HV variation.

The PMT gain variation as a function of HV is measured
	prior to running for all PMT's used in the calorimeter, 
	and is parameterized as:
\begin{eqnarray}\label{eq:hv-pmt-gain}
g_{PMT}=aV^{\alpha},
\end{eqnarray}
	where $g_{PMT}$ is the gain of the given PMT, $V$ is the
	HV in units of volts, and $a$ and $\alpha$ are the 
	fit parameters.
The exponents $\alpha$ are determined for each PMT, and 
	a typical value of $\alpha$ is $\sim6.8$.

The NuTeV experiment implemented a slow monitoring system 
	that monitored PMT HV values 
	as described in Section~\ref{ss:slow-mon}.
Six LeCroy 1440 HV mainframes supply 
	high voltage to the calorimeter PMT's.
The slow monitoring system read out 1 HV channel per second and
	completely cycled through all HV channels in 5 minutes.
The entire record in the database is written to the 
	neutrino data tape once every beam 
	cycle as the last record in the given cycle.
The readout resolution of the NuTeV HV slow monitoring 
\begin{figure}[tbp]
\begin{center}
\epsfxsize=0.45\textwidth
\epsfbox[30 150 540 670]{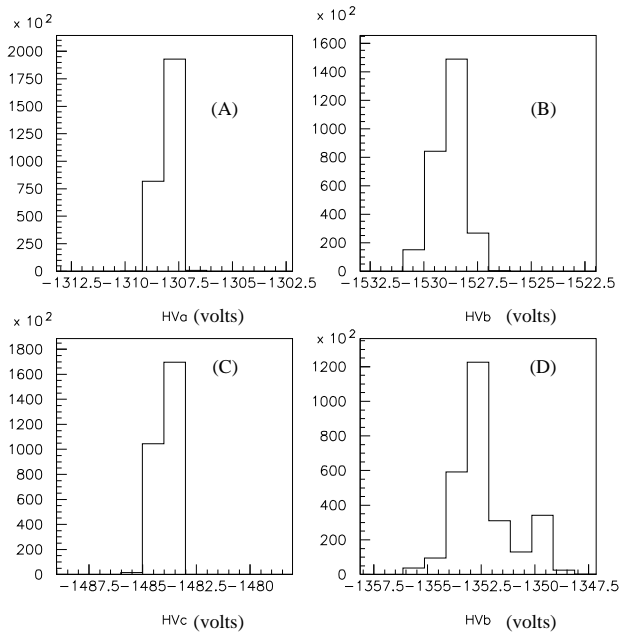}
\end{center}
\caption[]{Typical HV readout values for four randomly selected PMT's.
As one can observe, the HV readout does not vary more than 2-5~V.}
\label{fig:typical_hv}
\end{figure}
	system is $\sim1$~V; the PMT's are typically set at 
	between -1400~V and -1500~V.

A study based on a total of 280,000 measurements of each individual 
	HV read-back, taken over the entire run period, reveals 
	that the typical variation of each PMT HV readout is within 2~V 
	and the RMS of the distribution is typically less than 0.5~V.
Figure~\ref{fig:typical_hv} shows typical HV readout values of four 
	randomly selected PMT's for all the readings throughout the
	entire run.

Despite the fact that we expect very small corrections due to 
	HV variations based on the HV readout measurements, we 
	correct for HV for the calibration beam due 
	to the fact that calibration runs are typically 
	localized in time while the gain correction
	factors are averaged over a longer time period.

HV corrections for each counter are done using the measured
	parameters in Eq.~\ref{eq:hv-pmt-gain}, relative to 
	the counter gain correction factor at the center 
	of the counter, averaged over the muon response map period.
The relative gain correction factor is computed using 
	Eq.~\ref{eq:hv-pmt-gain} on an event-by-event basis 
	and normalizing the gain at the given HV readout to that 
	at the average HV readout of the PMT in the given run.
The relative correction factor for counter $i$, $f_{HV}^{i}$, is 
	defined as:
\begin{eqnarray}\label{eq:hv-cor-fac}
f_{HV}^{i}=\frac{ \sum_{j=1}^{4}PH_{ij}(<V_{ij}>/V_{ij})^{\alpha_{j}}} 
{ \sum_{j=1}^{4}PH_{ij}},
\end{eqnarray}
where $<V_{ij}>$ is the average HV readout value of PMT $j$ of the 
	counter $i$ for the given run, $V_{ij}$ is the HV readout for
	the given event, $\alpha_{j}$ is the exponent 
	in Eq.~\ref{eq:hv-pmt-gain} for PMT $j$,
	and $PH_{ij}$ is the individual pulse height from the PMT $j$.
As we expect, the typical overall size of this relative 
	correction factor is on the order of 0.1\% or less.

%
\section{Measurement of Muon Energy Loss in the Calorimeter and Comparison
with GEANT}
The toroid spectrometer is located downstream of the 690-ton NuTeV
        calorimeter. 
For an accurate measurement of each muon's momentum,
        the energy lost by the muon 
	in the calorimeter ($\Delta E$) has to be included.
A precise measurement of $\Delta E$ is also necessary for the 
	calibration of the toroid using test beam muons. 
Knowledge of muon energy depositions is also needed for the 
	hadronic energy measurement, since muons, originating 
	in $\nu_{\mu}$ charged current interactions,
        contribute to hadronic shower pulse heights.

A minimum-ionizing particle passing through the detector 
        leaves a characteristic energy deposit in each of 
        the scintillation counters. 
The energy loss of a muon traversing the calorimeter changes with energy.
For high energy muons the contribution to the muon energy loss 
	from bremsstrahlung, electron-positron pair production, and 
        nuclear interactions increases.
\begin{table}[tbp]
\begin{tabular}{|l|l|}\hline
Process&Value\\\hline
Rayleigh Scattering& ON (IRAYL=1)\\
$\gamma$-Induced Fission& ON (PFIS=1)\\
$\delta$-ray Generation above& DCUTE = 100 keV\\
Restricted Landau below& DCUTE = 100 keV\\
Direct Pair Production& ON (PPCUTM = 2.04 MeV)\\
Bremsstrahlung Tracking& (BCUTE = 100 keV)\\
 & (BCUTM = 100 keV)\\
Other Particles&CUTGAM=100 keV\\
 &CUTELE = 100 keV\\
 &CUTNEU = 100 keV\\
 &CUTHAD = 100 keV\\
 &CUTMUO = 1 MeV\\
\hline\hline
\end{tabular}
\caption{Parameters with changed values from their default values 
        in GEANT V3.215.}
\label{tab:geant}
\end{table}
These processes may yield larger electromagnetic showers than 
        would be true for a strictly ``minimum-ionizing" particle.  
A coarse sampling calorimeter, such as NuTeV's, is strongly non-linear 
        in energy between a fraction of an MeV to a few GeV.
Hence the conversion of the light yield induced by a 
        muon passing through the counters to measured muon 
        energy loss in the calorimeter requires a differentiation 
        between lower and higher energy processes. 
We use a GEANT-based simulation of the detector (McNuTeV) to determine 
	the best pulse height to GeV conversion method or  
        ``reconstruction algorithm'' for the total energy lost 
	by the muon, $\Delta E$.  
Tests of both the GEANT simulation and the 
	reconstruction algorithm are described in the following sections.  
\subsection{Counter Pulse Height Simulation using GEANT}
The NuTeV  calorimeter simulation  segments the calorimeter into
        six identical carts, each of which consists of seven 
        unit calorimeter layers described in
        Section~\ref{ss:detector}.
The steel, water, drift chamber gas, lucite, mylar, polythene, 
        air, copper, and G10 are specified as separate GEANT 
        volumes building the layers with sizes and configurations closely 
        matching the physical detector. 
We find that very detailed modeling of the detector is necessary to
        achieve good agreement between calibration beam data and 
        the GEANT simulation of muon responses.

We use version 3.215 of GEANT and set the physics control
        variables to their default values, with the exceptions listed
        in Table~\ref{tab:geant}.
The energy deposited in scintillation counters 
        follows an avalanche in our 10-stage model of PMT's
        and is smeared statistically at each step.  
The number of  photoelectrons used in the smearing is
        tuned to match the widths of muon $dE/dx$ deposition
        in the data.
Pedestals, gains, and the digitization of $LOW$ and $HIGH$
        channels of electronics are also simulated.
        Observed pulse heights, both for data and for the simulation, are
        expressed in units of MIP's, where 1 MIP is defined as a 
        truncated mean of the energy loss of 77~GeV muon 
        (see Section~\ref{ss:gain_cor}).
The resulting GEANT events are passed through the same analysis chain as 
        the actual data events.
\subsection{Data/GEANT Comparisons for Muons}
Muon calibration beam data are taken throughout the 1996-97 NuTeV run  
        totaling approximately 250 10000-event data sets. Most of that
        data are 50~GeV muon sets used for measuring the magnetic
        field of the toroid. 
Another subset, also used in this study, consists of 
        runs with muon energies spanning from 12.5 to 190~GeV.
GEANT samples are generated with the energies, momenta, and 
	positions at the entrance to the calorimeter matching the 
        calibration beam data samples. 
In the comparisons, 
        cuts are applied to the calibration beam muons to assure 
        that the momentum measured in the test beam spectrometer 
        and the $x$- and $y$-vertex positions are reconstructed 
        within $\pm$3 standard deviations around the mean value. 

\begin{figure}[tbp]
\begin{center}
\epsfxsize=0.45\textwidth
\epsfbox[40 150 525 650]{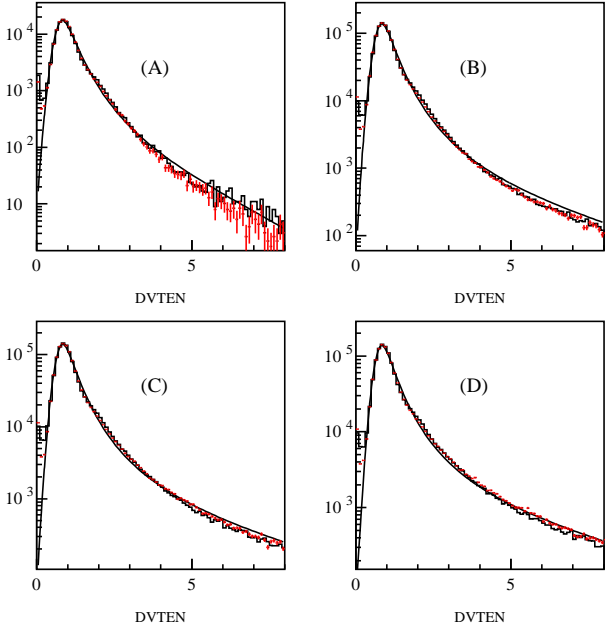}
\end{center}
\vspace{-0.1in}
\caption{Counter response to muons in calibration beam 
(histogram) and Monte Carlo (crosses) for muon of energies of 15 (A), 50 (B), 100 (C), and 166~GeV (D). 
DVTEN (HIGH/LOW channel depending on saturation) are
 muon map corrected and measured 
in MIP's. All 84 counters contribute to the distributions.  
Solid lines represent 5-parameter asymmetric Gaussian fits 
        (Eq.~\ref{eq:asym-gaussian}) to the distributions. }
\label{fig:dvten}
\end{figure}
\begin{figure}[tbph]
\begin{center}
\epsfxsize=0.45\textwidth
\epsfbox[40 150 525 650]{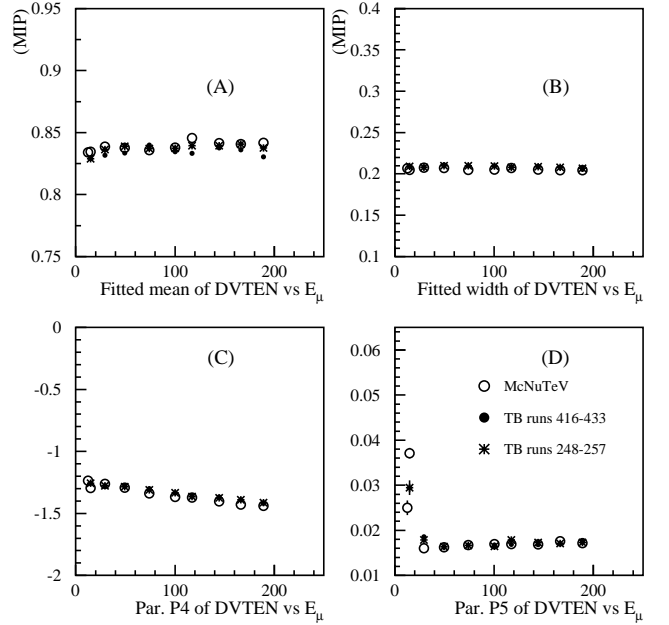}
\end{center}
\vspace{-0.1in}
\caption{Fit parameters (A, B: mean and RMS of the Gaussian part,
        C and D: asymmetric tail, Eq.~\ref{eq:asym-gaussian})
        to DVTEN distributions for Monte Carlo (open circles) and calibration 
        beam muons plotted versus incident muon energy (${\rm E_{\mu}}$).}
\label{fig:fit_dvten}
\end{figure}
Figure~\ref{fig:dvten} illustrates the  detector response
        to calibration 
        beam data (histogram) for muons of energies of 15, 50, 
        100, and 166~GeV. 
The GEANT simulation is marked by crosses.
DVTEN shown in the plots is the pulse height of a counter 
        measured in units of MIP's, after application of the 
	position dependent gain correction discussed in 
	Section~\ref{ss:muon-map}. 
It uses the $HIGH$ channel of electronics until its ADC saturation
	(1900 ADC counts), and the $LOW$ channel readout above that.
The DVTEN distributions in muon energy bins are fitted with a 
        five-parameter asymmetric Gaussian fit, ${\cal F}$, 
        in which the width of the Gaussian runs on one side, 
        varying with the $x$-axis: 
\begin{eqnarray}\label{eq:asym-gaussian}
{\cal F} &=& \frac{P_{3} e^{-(x- P_{2})^2/2\sigma^2}}
{\left| P_{1}~P_{3} \right| } \\
 &~& \nonumber \\
\sigma &=& P_{3}{\rm max}(1,(1 -(P_{4}+xP_{5})(x-P_{2})))
\end{eqnarray}

\begin{figure}[tbp]
\begin{center}
\epsfxsize=0.45\textwidth
\epsfbox[40 150 525 650]{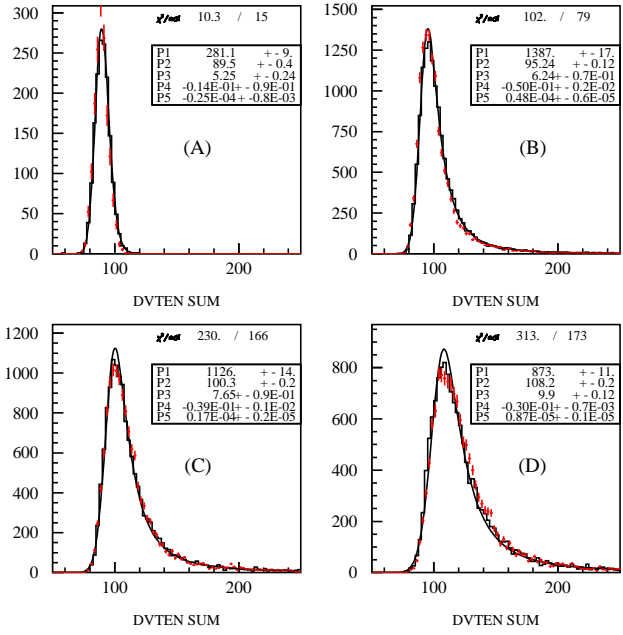}
\end{center}
\caption{Counter response to muons summed over 84 counters
 in calibration beam 
(histogram) and Monte Carlo (crosses)
for muon of energies of 15 (A), 50 (B), 100 (C), and 166~GeV (D).}
\label{fig:dvten_sum}
\end{figure}

\begin{figure}[tbph]
\vspace{-0.1in}
\begin{center}
\epsfxsize=0.45\textwidth
\epsfbox[40 150 525 650]{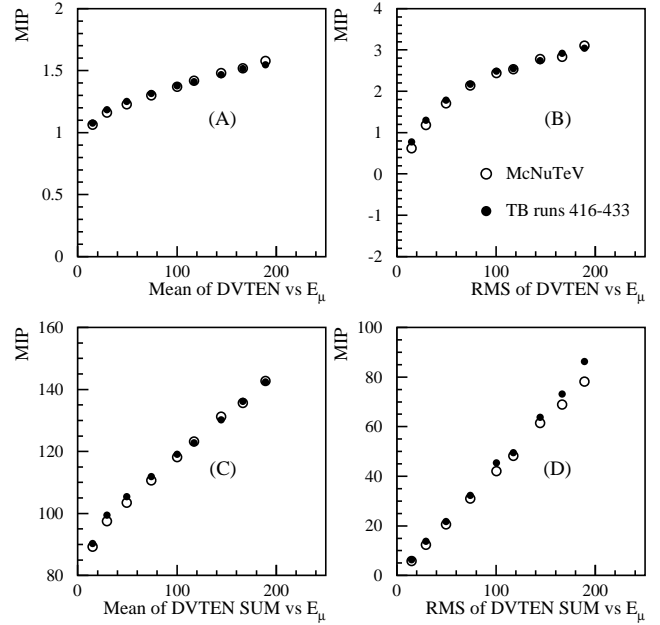}
\end{center}
\caption{Mean values of DVTEN (A) and DVTEN SUM over 84 counters (C) 
        distributions for Monte Carlo (open circles) and 
        calibration beam muons (solid circles) as well as RMS values of these
        distributions.}
\label{fig:dvten_abs}
\end{figure}

Figure~\ref{fig:fit_dvten} gives the values
        of the four  parameters of these fits (peak of the 
        Gaussian portion of the distribution ${\rm P_{2}}$, its 
        width ${\rm P_{3}}$, and, ${\rm P_{4}}$ and 
        ${\rm P_{5}}$, two parameters describing the
        asymmetric tail) as a function of muon energy.
The Monte Carlo is represented by open circles, while 
        the muon data sets by solid circles and stars.
The calibration beam data show that the most probable value of 
        muon energy loss in a counter is independent of muon energy
        in the 20-200~GeV range (see Figure~\ref{fig:fit_dvten}).  
This is to be contrasted with almost linear increase
        with muon energy of the mean energy loss in the counter 
        (also Figure~\ref{fig:dvten_abs}).
Figure~\ref{fig:dvten_sum} shows the pulse heights
        summed over 84 counters traversed by muons 
\begin{figure}[tbph]
\begin{center}
\epsfxsize=0.45\textwidth
\epsfbox[40 150 525 650]{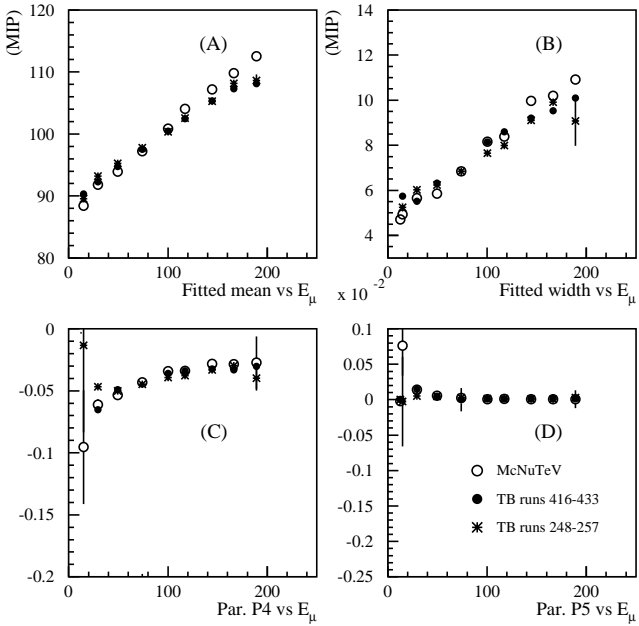}
\end{center}
\caption{Fit parameters (A, B: mean and RMS of the Gaussian part,
        C and D: asymmetric tail, Eq.~\ref{eq:asym-gaussian})
        to DVTEN SUM distributions for Monte Carlo (open circles) and calibration 
        beam muons plotted versus incident muon energy (${\rm E_{\mu}}$).
DVTEN SUM is the sum of the DVTEN over 84 counters.}
\label{fig:fit_dvten_sum}
\end{figure}
        (DVTEN SUM distributions).  
This summed plot would magnify small discrepancies (e.g., in the tails)
	in comparison of calibration data and the Monte Carlo,
 	but GEANT still describes the data well. 
A summary plot  containing fit parameters to DVTEN SUM histograms for
        all available muon energy points is shown in 
        Figure~\ref{fig:fit_dvten_sum}.
The mean and RMS values (not from fits, but from the histogram 
        statistics) for the DVTEN and DVTEN SUM distributions
        are plotted in Figure~\ref{fig:dvten_abs}.
From these calibration beam data-GEANT
        comparisons, we conclude that both the low and high energy 
        components of the muon energy loss in the 
        calorimeter are 
        well modeled in our Monte Carlo 
        over the full scale of such depositions.
\begin{figure}[tbph]
\vspace{-0.3in}
\begin{center}
\epsfxsize=0.45\textwidth
\epsfbox[27 150 525 650]{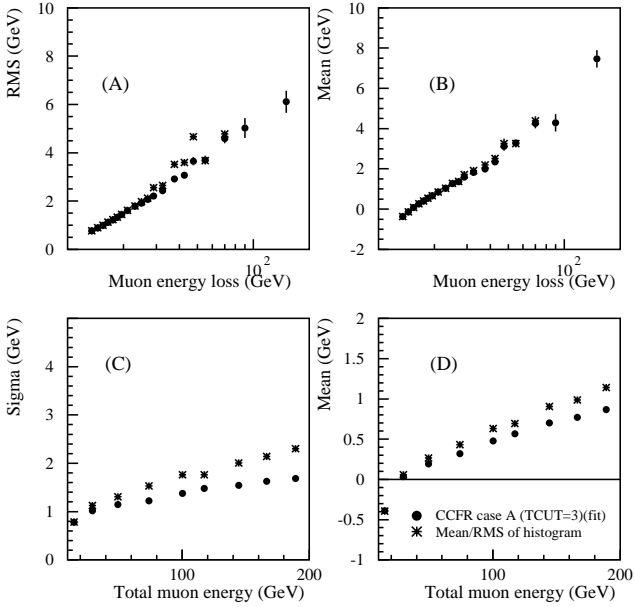}
\end{center}
\caption{Method $\alpha$: The width (A) and mean (B) from Gaussian fit 
        (solid circle) and histogram statistics (stars) of the 
        difference between ``true'' $\Delta E$ and reconstructed $\Delta E$ 
        distributions in muon energy loss bins.  C and D: the same 
        in muon momentum bins.}
\label{fig:ccfrdiff}
\end{figure}
Modeling of GEANT muon energy loss in the steel and remaining absorber
	materials can be checked using so-called range-out muons. Those are
	low energy calibration beam muons that stop in the calorimeter.
Our measurements give the mean length of 75.7$\pm$0.3 counters
 	for a 12.5~GeV muon in the data versus 74.7$\pm$0.09 
	in the simulation.
The agreement between data and Monte Carlo in the mean length is only
	within about three standard deviations.
One possible cause of this disagreement is the lack of any correlated
counter noise in the Monte Carlo.

\subsection{Study of Muon Energy Loss Reconstruction Algorithm Using GEANT}
\begin{figure}[tbp]
\begin{center}
\epsfxsize=0.45\textwidth
\epsfbox[30 150 520 650]{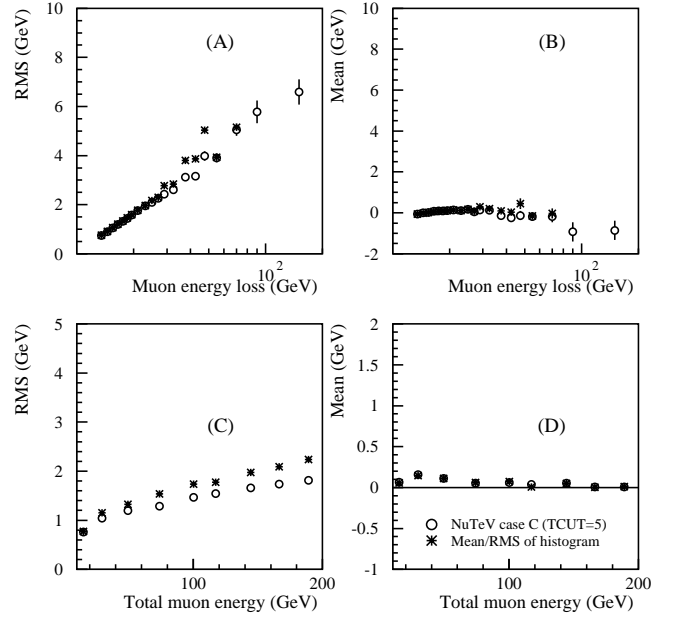}
\end{center}
\caption{Method $\gamma$.
The fitted widths (A) and means (B) of the difference between true $\Delta E$ and
 reconstructed $\Delta E$ distributions in muon energy loss bins. 
(C, D): the same in muon momentum bins.}
\label{fig:t5diff}
\end{figure}

\begin{figure}[tbph]
\begin{center}
\epsfxsize=0.42\textwidth
\epsfbox{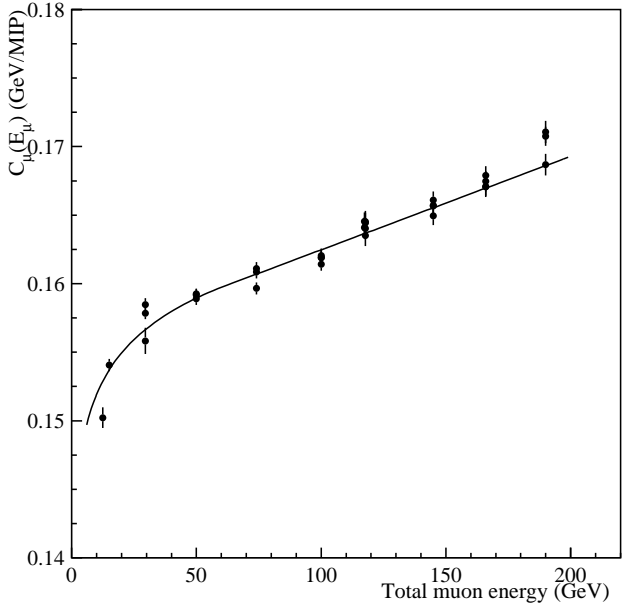}
\end{center}
\vspace{-0.6in}
\caption{Energy dependence of calibration beam muon depositions 
of pulse heights below 5 MIP's, over 84 counters,
  shown in units of GeV/MIP and the
parameterization $C_{\mu}(E_{\mu})$.}
\label{fig:cmu}
\end{figure}

The goal of this study is to find an optimal algorithm to determine the
	muon energy loss in the calorimeter, $\Delta E$, from the observed 
        muon pulse height.
We reconstruct $\Delta E$ and 
        compare it to the ``true'' $\Delta E$ known from GEANT 
        on an event-by-event basis in the range of 15--190~GeV.  
We define TCUT as the counter pulse height at which we 
        switch from applying $C_{\mu}$ (``low energy'') 
        to $C_e$ (``high energy'') conversion from MIP's to GeV,  
        under the assumption that sufficiently high pulse heights 
	arise from electromagnetic processes sampled over 
	several counters.
Three different reconstruction schemes for $\Delta E$ are studied:

\begin{description}
\item[Method $\alpha$:] A model of two conversion constants $C_{\mu}$ and $C_{e}$ and TCUT of 3 MIP's (this is a scheme used in our predecessor experiment 
CCFR, where $C_{e}$ is determined from electron calibration beam data).
\item[Method $\beta$:] A ``one-function model'', where one function 
        $C_{\mu}(E_{\mu})$ is used to account for ionization and 
        the increase of the radiative component of $dE/dx$ with energy. 
\item[Method $\gamma$:] A model of  conversion function $C_{\mu}(E_{\mu})$
        varying 
        with muon 
        energy $E_{\mu}$, applied 
        below TCUT of 5 MIP's, and a constant conversion
        $C_{e}$ above that TCUT.
\end{description}
As an illustration, we show the widths and the means 
        of the difference between the ``true'' 
        and reconstructed $\Delta E$
        distributions in either muon energy bins or 
        in the bins of the muon true energy loss for Method $\alpha$ 
          in Figure~\ref{fig:ccfrdiff} and Method $\gamma$ 
        in Figure~\ref{fig:t5diff}.
Both the fitted mean and sigmas of Gaussian fits to these distributions 
        (solid circles) and the average values and RMS (stars) of these 
        distributions are plotted. 
Notice that Method $\alpha$ underestimates $\Delta E$ for high
        energy electromagnetic depositions if the lower end 
        of the energy loss spectrum is set to match GEANT's ``true'' 
        $\Delta E$ (Figure~\ref{fig:ccfrdiff}(B)). 
Similarly for Method $\beta$ (not shown) -- no $C_{\mu}(E_{\mu})$
        can be found that describes the conversion from MIP's to GeV for
        both the most probable and the mean $dE/dx$ at the low and high
        ends of muon energy spectrum at the same time.
In Method $\gamma$, where the variation in low-energy radiative 
	depositions with 
        muon energy is accounted for by variation of the
        conversion function $C_{\mu}(E_{\mu})$, we find the best 
        match of the ``true'' $\Delta E$ for all muon
        energies and $\Delta E$ values (Figure~\ref{fig:t5diff}).
The function $C_{\mu}(E_{\mu})$, based on the best calibration 
	beam-to-GEANT match, is shown in Figure~\ref{fig:cmu}.

\begin{figure}[tpbh]
\begin{center}
\epsfxsize=0.42\textwidth
\epsfbox{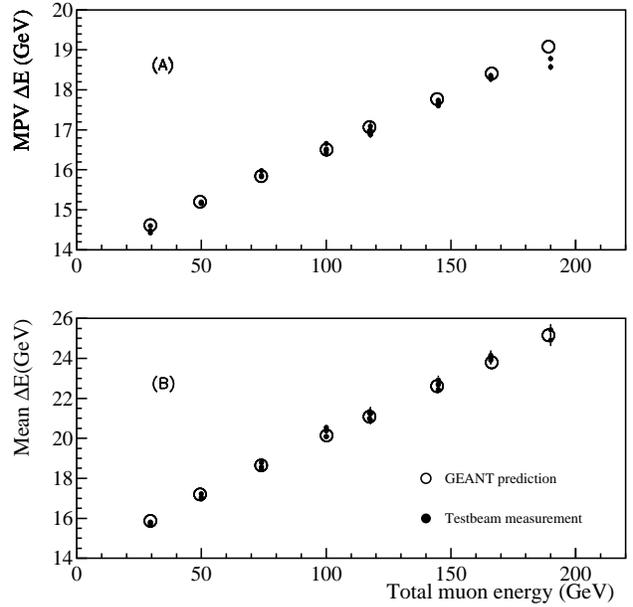}
\end{center}
\vspace{-0.6in}
\caption{Most probable (A) and mean value (B) of total muon energy loss 
in NuTeV calorimeter.
Comparison of GEANT prediction (open
circle) and calibration beam measurement (solid circles) versus muon energy.}
\label{fig:compare}
\end{figure}

\begin{figure}[t]
\begin{center}
\epsfxsize=0.42\textwidth
\epsfbox{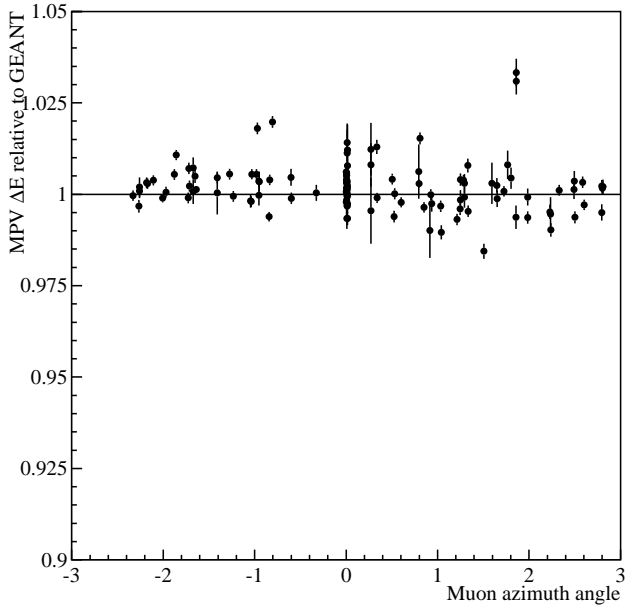}
\end{center}
\vspace{-0.65in}
\caption{Ratio of the most probable values of $\Delta E$ for 50~GeV 
        calibration beam muons to GEANT prediction (15.2~GeV) 
        as a function of muon azimuthal angle.}
\label{fig:muon_map}
\end{figure}

Figure~\ref{fig:compare} illustrates the
	total muon energy loss over the length of NuTeV calorimeter 
        in terms of mean and the most probable value, where
        the latter is defined as the result of a fit of asymmetric 
        Gaussian function ${\cal F}$ (parameter P2) to the 
        $\Delta E$ distribution.
Figure~\ref{fig:muon_map} gives a ratio of the most probable $\Delta E$ for 
        50~GeV calibration beam muons, traversing the NuTeV calorimeter 
        at different angles and transverse positions, to a 
        nominal 15.2~GeV GEANT prediction for their energy loss.  
The ratio is plotted as a function of muon azimuthal angle, $\phi$, at 
        the most upstream surface of the detector. 
As can be seen in this figure, we reconstruct muon $\Delta E$ to within 
        $\pm0.7\%$,  independent of $\phi$ (or position in the counter). 
This is an important verification of the muon map correction
        and counter gain stability over time.
The counter pulse heights for these calibration beam muon samples
        are corrected by gains that vary by as much as an 
	order of magnitude, depending on their transverse 
	vertex and pathway through the calorimeter.

%
\section{Shower Energy Definition}\label{ss:ehad}
The definition of the shower energy used in the hadron energy 
	calibration is:
\begin{eqnarray}\label{eq:shower_energy}
E_{shower}=C_{\pi}\left[ \sum_{i=1}^{place}PH_{i}
+\sum_{i=place}^{place+19}h_iPH_{i} \right] , 
\end{eqnarray}
where $i$ is the counter number; $PH_{i}$ is the pulse 
	height normalized for the muon gain as described in 
	Section~\ref{ss:gain_cor} in units of MIP in 
	counter $i$; $h_i$ is the hadron/muon gain ratio 
	(described in Section~\ref{ss:hadgainbal}); and
	$C_{\pi}$ is the hadron calibration constant. 
The variable $place$ is the counter where the hadron shower started to 
	develop, and is determined by an algorithm designed to locate where a 
	neutrino interaction begins.  
$place$ is defined as the upstream of two consecutive counters 
	which have more than a certain number
	of MIP's, where that number depends on the total energy of the 
	hadron shower and is at least four.  
Upstream of $place$ the hadron is treated as a minimum ionizing particle.  
In contrast, electron showers always start at the first counter.  
For the electron energy measurements described later only the seven 
	most upstream counters are used, and the hadron/muon gain 
	ratio is applied to each counter's pulse height.  

In order to determine the hadron energy calibration constant which, in our
	definition, is the GeV-to-MIP conversion factor ($C_{\pi}$), 
	it is important to define the energy variable to 
	contain the entire shower.
On the contrary, for a precision measurement, one does not want to 
	include too many counters in the sum because adding 
	more counters than necessary would introduce 
	noise into the system due to pedestal fluctuations 
	in the counters.
Finally, the algorithm should be as close as possible to that used in 
	analyzing the neutrino data, which also sums over a certain 
	number of counters 
	following $place$.  
We perform a study to optimize the number of counters over which to sum the
	pulse heights.
Using hadron beams over the energy range between
	10~GeV and 190~GeV, we determine that summing over 20 counters 
	beginning from the most upstream counter is optimal for 
	calibration purposes.

\begin{figure}[tbp]
\begin{center}
\centerline{\psfig{figure=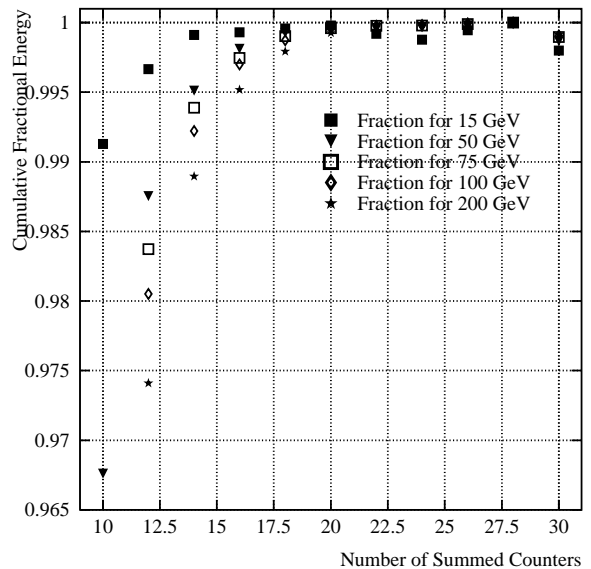,width=3.5in,height=3.5in}}
\end{center}
\vskip -0.4in
\caption[]{Cumulative fractional shower energy as a function of
the added number of counters.   Note that for all energies, adding
20 counters provides full longitudinal shower containment.}
\label{fig:shower_contain}
\end{figure}
Figure~\ref{fig:shower_contain} shows the cumulative fractional energy 
	as a function of the number of summed counters for various 
	hadron beam energies.
Since we summed the pulse heights of at least 20 
	counters dependent on the measured hadron energy, we 
	introduce an energy dependent noise level.
This noise level depends on beam energy because the shower
	penetration depth depends on beam energy.
The number of counters without actual shower 
	energy increases with decreasing beam energy and 
	the noise level gets more prominent for low 
	energy beams. 
Therefore, the low energy calibration has a larger 
	contribution from this noise effect.
However, since hadronic energy resolution is worse at low 
	energies, this noise is less important.

%
\section{Hadron Energy Response and Resolution}
\label{sec:hadron} 
\begin{figure}[tbp] 
\begin{center}
\centerline{\psfig{figure=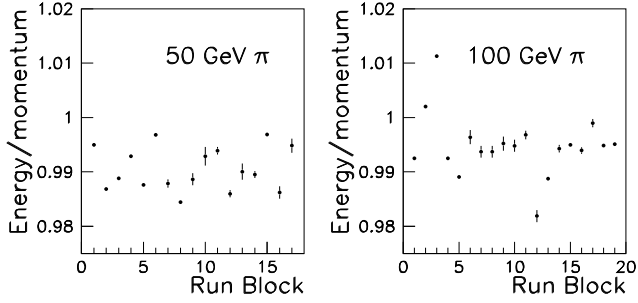,width=.475\textwidth}}
\end{center}
\vspace{-0.3in}
\caption[]{Remaining time dependence of the ratio of reconstructed
calorimeter energy divided by measured beam momentum.  The left 
graph is for 50~GeV runs, and the right is for 100~GeV runs.  
The horizontal axis is time in units of blocks, where each 
run block corresponds to a period of about 2 weeks.  
The responses are obtained after all the time-dependent corrections are 
	applied to the data, but before the final energy scale 
is set.}  
\label{fig:timedep} 
\end{figure}

The simplest test of the muon calibration technique described above is 
	the time dependence of a particular calibration beam setting.  
Figure~\ref{fig:timedep} shows the time dependence of both 
	50~GeV hadron and 100~GeV hadron runs
	that are taken periodically during the course of the 
	experiment.  
The RMS of the ratio between reconstructed calorimeter energy 
and beam momentum is 0.4-0.5\%, and is due to the 
	statistical uncertainty in the muon maps themselves, 
	as well as the electronics gain coefficients.  

\begin{figure}[tbph]
\begin{center}
\centerline{\psfig{figure=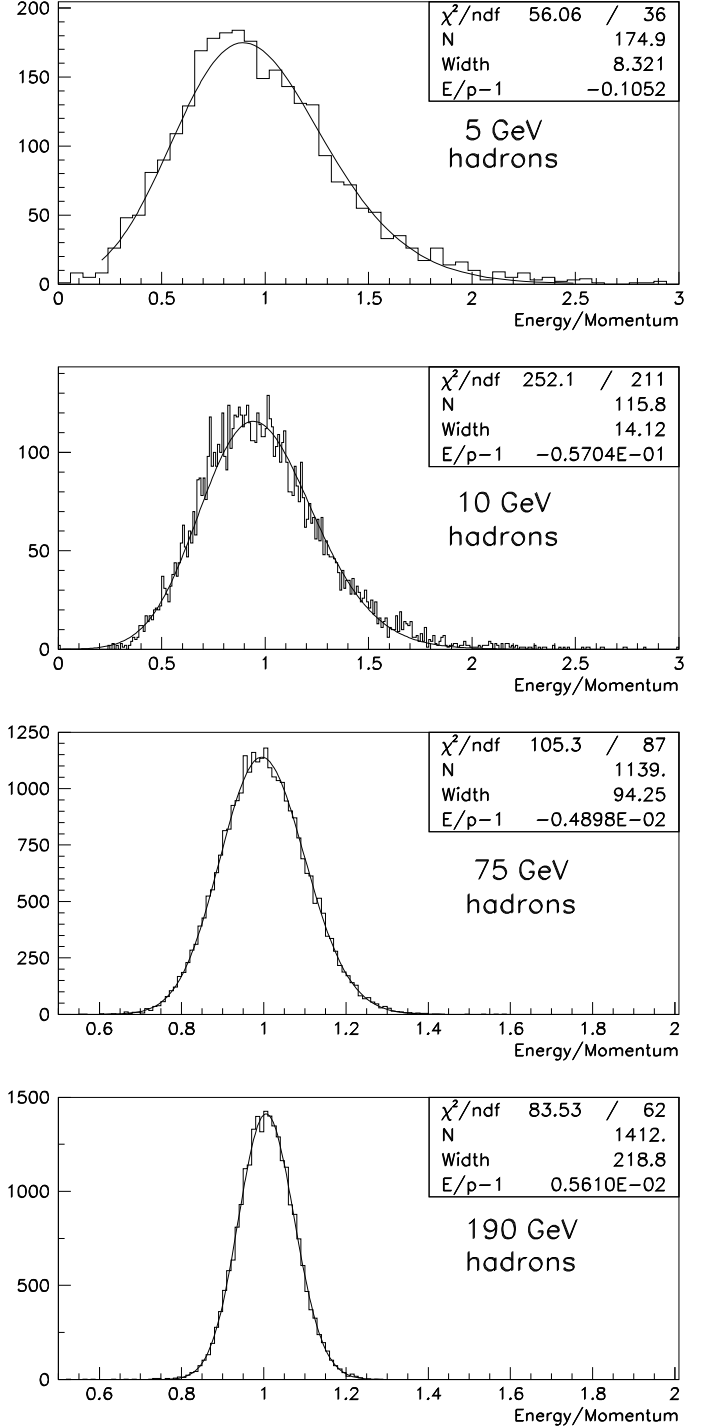,width=.5\textwidth}}
\end{center}
\vspace{-0.3in}
\caption[]{ Poisson fits to the calorimeter energy divided by
 momentum distributions for four different energies:  
5~GeV, 10~GeV, 75~GeV, and 190~GeV.}  
\label{fig:lohieng}
\end{figure}

\begin{figure}[bp]
\begin{center}
\centerline{\psfig{figure=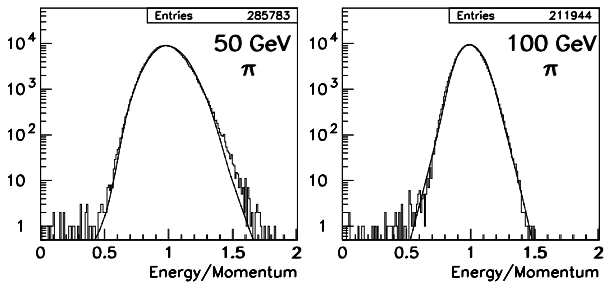,width=.45\textwidth}}
\end{center}
\vspace{-0.3in}
\caption[]{The distributions of calorimeter responses to 50 and 100~GeV
hadrons with very high statistics.   The data follow the Poisson-like 
shape over several decades.  }
\label{fig:hi-stat}
\end{figure}

The calorimeter response to a monochromatic beam of hadrons
        can be characterized by a function similar to a Poisson
        distribution.
This is because the energy reconstructed by the calorimeter is
        proportional to the number of shower particles
        produced by the incident hadron.
The statistical fluctuation of the number of electromagnetic
        particles in the shower causes the response to
        look Poisson-like at low energies, and to
        become Gaussian at high energies.   This can be seen in
figure \ref{fig:lohieng}, where the 5~GeV data are much less symmetric
than the 190~GeV data around the peak of the Energy/momentum distribution.
Figure~\ref{fig:hi-stat} shows that the Poisson-like function, defined
below, describes the data over several decades.
Fluctuations in where the primary hadron interaction occurs can also
contribute to the asymmetric shape of this distribution, but again these
fluctuations have a negligible effect at high energies.

The Poisson distribution is normally written as 
\begin{eqnarray}\label{eq:poisson}
P(N,\mu) = \frac{\mu^N e^{-N}}{N!}, 
\end{eqnarray}
where $P(N,\mu)$ is the probability of 
	seeing $N$ shower particles if $\mu$ are expected.   
The RMS of this distribution is $\sqrt{\mu}$ and the 
	mean is $\mu$.  
As $N$ gets large this approaches a simple Gaussian distribution.  
To remove any effects from variations in time of the 
	calibration beam momentum, the calorimeter energy is divided by the 
	reconstructed particle momentum on an event-by-event basis.  
This implies that the mean of the distribution is decoupled from the 
	width, but the fractional width (width divided by the mean) 
	remains $1/\sqrt{\mu}$.  
Generalizing Eq.~\ref{eq:poisson} to decouple the mean from 
	the width and expanding about the peak, we 
	can parameterize the Poisson distribution as follows
	(note that keeping only the first term in $F(x)$ gives 
	a Gaussian distribution):  
\begin{eqnarray} 
P(x) &=& A e^{-F(x)}, 
\end{eqnarray}
where
\begin{eqnarray} 
x&=&B(E/p - C),
\end{eqnarray}
and
\begin{eqnarray} 
F(x) &=&  \frac{1}{2} 
\left( B - \frac{1}{2} + \frac{1}{24B} \right) 
(\frac{x}{B}-1 )^2  \nonumber\\
& &+ \frac{1}{6} \left( B - \frac{1}{4} + \frac{1}{72B} \right) 
(\frac{x}{B}-1 )^3   \nonumber \\
& &- \frac{1}{48} \left( B - \frac{1}{6} \right) (\frac{x}{B}-1 )^4 ,
\end{eqnarray} 
where $E$ is the measured hadron energy, $p$ is the reconstructed 
	calibration beam momentum, 
	the peak $E/p-1$ of the distribution is $C$, 
	and the width of the distribution is $B$.  
The fractional width of the distribution is $1/\sqrt{B}$. 
At beam momenta of 30 GeV and above, this equation is very close to a 
	Gaussian distribution.  
Figure~\ref{fig:lohieng} shows fits to the above equation for four 
	different energies.  

\begin{figure}[bph]
\vspace{-.1in} 
\begin{center}
\centerline{\psfig{figure=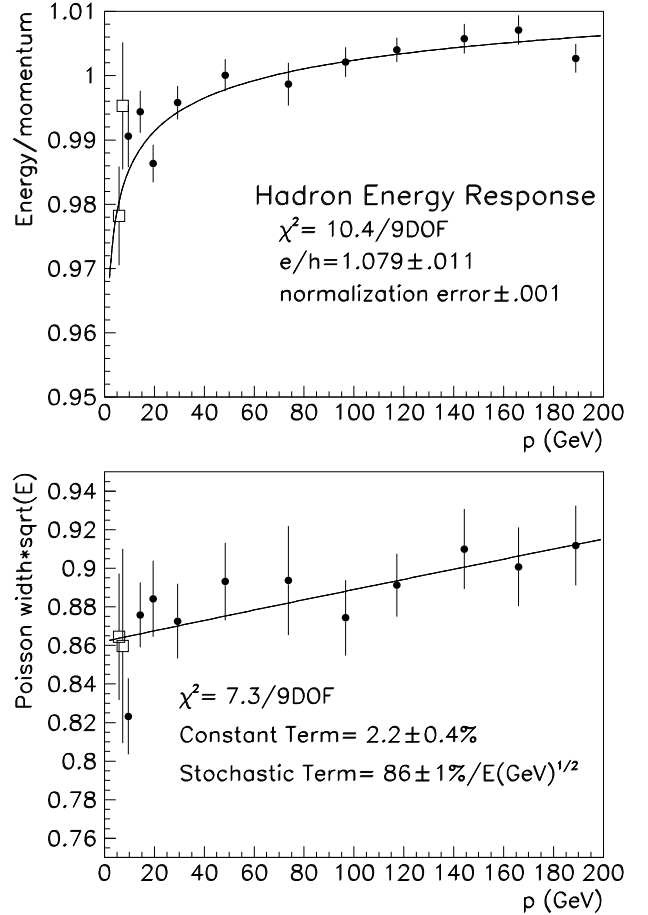,width=.45\textwidth}}
\end{center}
\vspace{-0.3in}
\caption[]{Hadron energy response versus reconstructed
test beam momentum and comparison with fit 
to Groom's parameterization for non-linearity, and 
Poisson width distribution versus energy with fit to 
$\sigma(E)/E = A \oplus \frac{B}{\sqrt{E}}$.  The open symbols 
are lower energy runs with slightly different cuts and are not 
used in the fits. }  
\label{fig:hadres} 
\end{figure}

The energy dependencies of the mean $E/p$ distribution and 
	the Poisson widths are shown in Figure~\ref{fig:hadres}.  
If the hadron calibration constant $C_\pi$, as defined in Section 
\ref{ss:ehad}, is set to $0.212$ after all corrections, the mean
energy response divided by the reconstructed test beam momentum 
at 75~GeV is $1.000\pm 0.001$.  
Note that the non-linearity of the calorimeter between 10~GeV and 
	190~GeV is only~$3\%$.  
This comes from the fact that electrons and hadrons have a very similar
	response, and so the electromagnetic component of the shower, which 
	changes as a function of energy, will not change the 
	reconstructed energy.  
The Poisson widths can be fitted to the standard form 
	$\sigma(E)/E = A \oplus \frac{B}{\sqrt{E}} \oplus \frac{C}{E}$, 
	where $A$ is a constant term coming from calibration uncertainties, 
	$B$ is the stochastic term from the sampling of the shower, 
	and $C$ is from noise due to pedestal fluctuations.  
The data show no evidence for a noise term and so $C$ is removed from the fit. 
The stochastic term is proportional to the square of the thickness of 
the sampling layer.  

\begin{figure}[tbp]
\vspace{-.35in} 
\begin{center}
\centerline{\psfig{figure=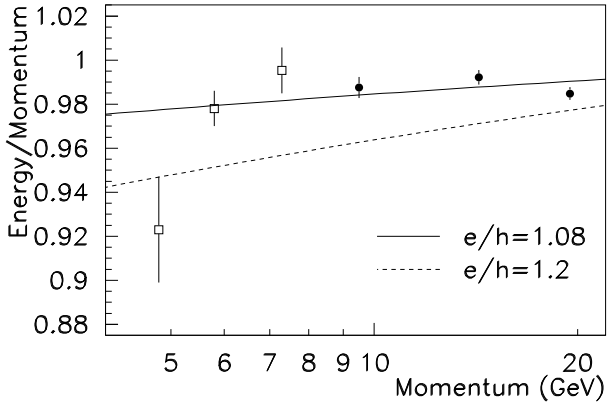,width=.5\textwidth}}  
\end{center}
\vspace{-0.3in}
\caption[]{Lowest energy hadron response versus reconstructed
test beam momentum and comparison with fit 
to Groom's parameterization for non-linearity. 
The open symbols are lower energy runs with slightly different cuts.} 
\label{fig:lowe_hadres} 
\end{figure}

The energy points below 10~GeV have to be measured using a 
	different energy algorithm, since the one that is designed 
	for neutrino interactions (requiring at least 
	4 MIP's in two consecutive counters) is biased for hadron 
	showers below 10~GeV.  
For these lowest points, the energy is determined
	simply by summing the most upstream twenty counters, and 
	is using the hadron gain coefficients for each counter.  
To remove electrons from the low energy  
	samples, cuts are made based on \u{C}erenkov counter 
	particle identification system in the beamline.  
To remove muons in the hadron beam, a loose cut is made on the most 
	upstream of three consecutive counters that have less than 
	0.25 MIP's in them.  
The latter cut removes events caused by muons in the hadron beam, but did not 
	remove events with secondary muons created in the hadron shower.  
Finally, because the lowest energy points have low statistics 
 	the means of the energy/momentum distributions are 
	plotted rather than the results of the Poisson fits.   
Figure \ref{fig:lowe_hadres} shows the nonlinearity of the NuTeV 
	calorimeter to low energy hadrons.  
For energies 5.9\,GeV and above, Groom's parameterization 
	(see section \ref{sec:hadronmc}) with $e/h=1.08$ (solid line) 
	agrees well with the data.  
The overall agreement with the parameterization is not improved 
	by changing $e/h$ from 1.08 in the hadron response curve, 
	as is also shown in figure \ref{fig:lowe_hadres}.

Finally, any additional position dependence not taken into account
	in the muon map procedure outlined earlier is studied using a 75~GeV 
	hadron beam aimed at different locations on the front face of the 
	calorimeter using the rotating dipole at the end of the momentum
	analyzing magnet chain in the spectrometer.  
For hadron showers that start more than 50~cm from the closest 
	edge of the detector, the energy reconstruction is constant 
	to better than a $0.5\%$ in the calorimeter response, when 
	normalized using the muon maps.  
By aiming the hadron beam as close to the edges as is safe, it is 
	determined that hadron shower leakage does not begin to affect energy
	reconstruction until the shower starts at 25~cm from the calorimeter
	edge.  

Table \ref{tab:systerrhad} lists systematic errors that 
	contribute to the uncertainty in overall hadron energy scale 
	of the calorimeter.  
It is clear from the list that the largest single systematic error is 
due to the statistical uncertainty in the hadron/muon gain ratio for 
the counters.  

\begin{table}[tbph] 
\begin{tabular}{|l|l|} 
\hline
 & Fractional \\
Source &  Uncertainty \\
\hline 
Hall Probe Readout & $0.03\%$ \\
(from shunt comparisons, see Figure~\ref{HP vs I}) & \\ 
\hline 
Magnetic Field Homogeneity & $0.03\%$ \\ 
(from Position-dependence measurements) & \\ 
\hline 
Beam Composition Corrections & $0.03\%$ \\ 
($100\%$ of effect above 30~GeV) & \\ 
\hline 
Transition Effect Uncertainties & $0.03\%$ \\
($10\%$ of effect on hadrons) & \\ 
\hline 
Temperature Corrections & $0.02\%$ \\ 
(10\% of effect) & \\ 
\hline 
High Voltage Correction & $<0.01\%$ \\ 
\hline 
Spectrometer Alignment & $0.1\%$ \\ 
\hline 
Uncertainty in hadron/muon gain ratios & $0.4\%$ \\ 
\hline
Fit Normalization Error & $0.1\%$ \\  
\hline
Statistical & $<0.01\%$ \\ 
\hline 
\hline 
Total & $0.43\%$ \\ 
\hline 
\end{tabular} 
\caption[]{ Table of contributions to uncertainty in overall hadron
energy scale.  }  
\label{tab:systerrhad} 
\end{table}

%
\section{Hadron Response Comparison to Monte Carlo}\label{sec:hadronmc} 
The task of reproducing the calorimeter attributes 
        in a GEANT-based Monte Carlo is a challenging one.  
This is demonstrated in Figure~\ref{fig:hadres}. 
Both the non-linearity and hadron energy 
	resolution of the calorimeter depend critically on the 
	difference in the calorimeter's response to hadrons 
	and to electrons ($e/\pi$).  
In order for a Monte Carlo to simulate hadrons correctly, it must 
	first correctly simulate the calorimeter's electron response, 
	and then have an accurate hadron shower model, 
	a thorough description of the geometry of the calorimeter, 
	and an accurate model for the way particles propagate in 
	the particular media that comprise the calorimeter.  
Section~\ref{sec:elecs} describes the calorimeter electron 
	energy response and the resolution in 
	detail; the studies show that the electron energy resolution 
	is well-modeled in the detailed GEANT simulation 
	of the calorimeter.  

\begin{figure}[tbp]
\begin{center}
\centerline{\psfig{figure=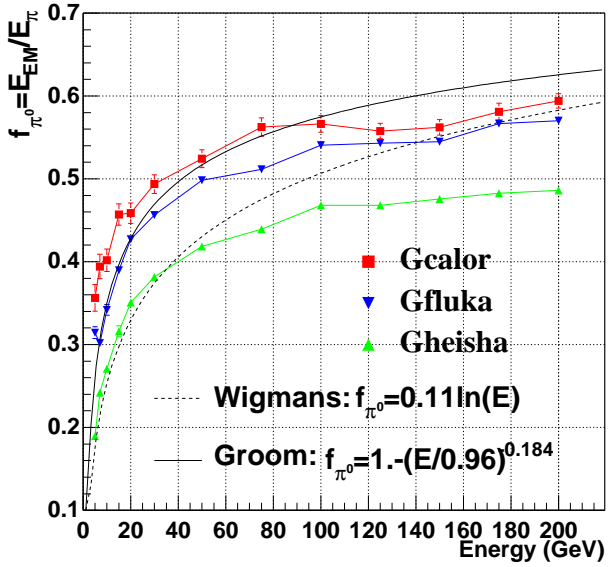,width=.475\textwidth}}
\end{center}
\vspace{-.3in} 
\caption[]{ Fraction of electromagnetic energy in a hadronic shower 
as a function of hadron energy for three different hadron shower 
generators:  GHEISHA, GFLUKA, and GCALOR.  
Fits of the simulation to both the Wigmans'~\cite{wigmans} and 
	Groom's~\cite{groom} parameterizations are also shown.}  
\label{fig:fpi0} 
\end{figure}

Once the electron to hadron response is measured in the data at 
	a particular energy, one can minimize the dependence on hadron 
	shower generators by only using them to predict the fraction of 
	$\pi^0$'s produced in a hadron shower as a function of energy,
	$f_{\pi^0}(E)$.  
Figure~\ref{fig:fpi0} shows three different 
	hadron generators' predictions for the fraction of electromagnetic
	energy deposited in a hadronic shower as a function of energy.  
Two parameterizations for this fraction are also shown; 
Wigmans' parameterization~\cite{wigmans} is 
	$f_{\pi^0} = 0.11\ln(E)$ and Groom's is
$f_{\pi^0} = 1-( \frac{E}{0.96~{\rm GeV}})^{-0.184}$.  

The reconstructed energy of a shower is defined as 
\begin{eqnarray}
E = E_{true} (h(1-f_{\pi^0}(E)) + e f_{\pi^0}(E)), 
\end{eqnarray}
where $h$ is the ratio of reconstructed to true energy
	for a ``pure'' hadron, and $e$ is that same ratio 
	for a ``pure'' electron.  
The non-linearity as a function of energy 
	can be expressed as the ratio of $R(E)/R(E_{ref})$, where 
	$R(E)= E/E_{true}$.  In other words, 
\begin{eqnarray}\label{non-linearity}
\rm\textstyle{non\,-\,linearity} = \frac{ 1-f_{\pi^0}(E) + \frac{e}{h}f_{\pi^0}(E)} { 1-f_{\pi^0}(E_{ref}) + \frac{e}{h}f_{\pi^0}(E_{ref})}. 
\end{eqnarray}

\begin{figure}[tbp]
\begin{center}
\centerline{\psfig{figure=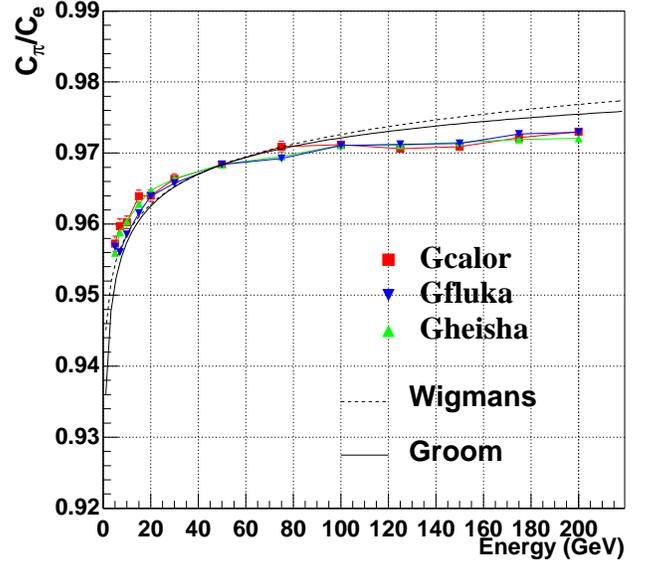,width=.475\textwidth}}
\end{center}
\vspace{-.3in} 
\caption[]{ Predicted non-linearity for three different hadron shower 
generators:  GHEISHA, GFLUKA, and GCALOR, where all three are 
required to have the same value of $E/E(true)$ at 50~GeV.  The prediction
for both the Wigmans and Groom parameterizations are also shown.}  
\label{fig:nonl} 
\end{figure}

By requiring the three generators shown in Figure~\ref{fig:fpi0} to 
	have the same $R(E)/R(E_{ref})$ at 50~GeV, one can construct the 
	expected non-linearity as a function of energy.  
This is shown in Figure~\ref{fig:nonl}.  
Although the fraction of electromagnetic energy 
	at a given energy varies among the different generators, the 
	predicted non-linearity is similar.  

By fitting the hadron energy response shown in Figure~\ref{fig:hadres},
	assuming the non-linearity predicted by Groom's parameterization, 
	one arrives at a value of $\frac{e}{h}$ of \hbox {$1.079\pm 0.011$}.  
Given that at 75~GeV the fraction of $\pi^0$'s is roughly $50\%$, 
	$\frac{e}{\pi}-1=0.5(\frac{e}{h}-1)$.  
Therefore, the electron to hadron response ratio is roughly 
	of \hbox {$1.035\pm 0.01$} at 75~GeV.    
This is in agreement with what is seen in the calibration beam
	comparisons between hadrons and electrons, as will be 
	described in Section~\ref{sec:elecs}.  
A similar conclusion has been reported by the CDF 
	collaboration~\cite{cdf} for their plug-upgrade hadron 
	calorimeter, which has a much larger non-linearity.  

As a side note, if we use the different hadron shower generators to predict
	the ratio of electron to hadron response at 75 GeV, they all give 
	different ratios.  
Since the difference between these responses 
	is also a large factor in determining the hadron energy resolution, 
	they all predict correspondingly different hadron energy 
	resolutions (the closer the electron to hadron response ratio 
	is to unity, the better the hadron energy resolution).  

GFLUKA~\cite{fluka} predicts the lowest ratio of 
	electron to hadron response, and also predicts a hadron
	energy resolution of $0.80/\sqrt{E(GeV)}$.  
GHEISHA~\cite{gheisha} predicts a ratio of electron to hadron 
	response of 1.15 at 75~GeV rather than 1.09, and 
	predicts a hadron energy resolution of $1.15/\sqrt{E(GeV)}$.  
Finally, GCALOR~\cite{gcalor} predicts that the ratio of electron to 
	hadron response is less than unity, rather than greater 
	than unity.  

All three generators are tested using identical GEANT 
	energy cutoff settings and identical geometry input.  
Although GHEISHA is native to the GEANT program, GFLUKA and GCALOR 
	are imperfect implementations of the original FLUKA and 
	CALOR programs and have been known to produce somewhat different 
	results than the original generators~\cite{atlas}.  

Finally, although the hadron non-linearity is now parameterized 
	and well-measured, the purpose of the NuTeV calorimeter 
	is to measure hadron showers generated by neutrino 
	interactions, not hadron showers generated 
	by a single charged hadron.  
To study any possible difference 
	the LUND Monte Carlo program is used to determine the first 
	set of particles produced from the hadron shower of 
	a neutrino interaction.  
Groom's parameterization is then used to calculate the electromagnetic 
	fraction of the charged hadrons which get produced.  
Although the charged hadrons have lower energy than the initial 
	total hadron energy, and as such would have a lower 
	electromagnetic fraction, there are also 
	neutral pions that are produced, which increase the electromagnetic
	fraction.  
The two effects cancel, keeping the 
	electromagnetic fraction as a function of total hadron energy 
	the same between neutrino-induced hadrons and single hadrons, 
	to a few per cent.  The resulting  
	effect on the neutrino-generated hadron non-linearity is 
 	negligible compared to the statistical error on the e/h fraction 
	itself, and the overall energy scale change is consistent with zero
	to better than $0.1\%$.

%
\section{Electron Energy Response and Resolution}\label{sec:elecs} 
The calorimeter response to electrons can be measured using the 
	calibration beam when set to the electron mode, as described in 
	Section~\ref{sec:beamselect}.  
Although there is a large contamination of muons in the electron 
	running, this is easily removed from the 
	data sample by looking at the most downstream counter 
	and by selecting events with more than 
	one minimum ionizing particle in them.  
Since electrons penetrate no more than a few cm of steel, 
	most of the energy is deposited in the first three counters, 
	so calibrating the detector response to electrons is 
	extremely dependent on the muon maps of those three 
	counters, and has larger systematic uncertainties 
	due to statistical errors in the muon maps.  

\begin{figure}[bp]  
\begin{center} 
\centerline{\psfig{figure=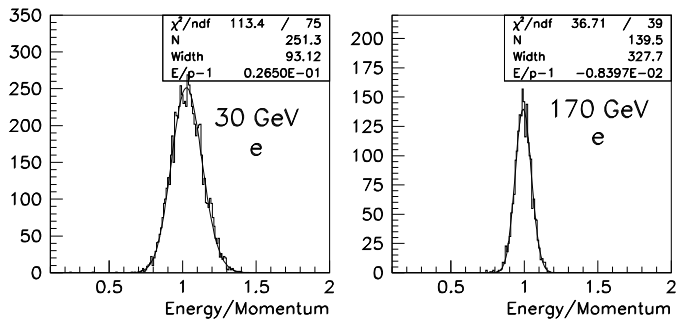,width=.5\textwidth}}
\end{center}
\vspace{-.3in} 
\caption[]{Calorimeter energy divided by test beam reconstructed
momentum distributions for 30~GeV and 170~GeV electrons, and the 
results from the Poisson fit to the distribution.} 
\label{fig:eleeop} 
\end{figure} 
\begin{figure}[tbp]
\begin{center} 
\centerline{\psfig{figure=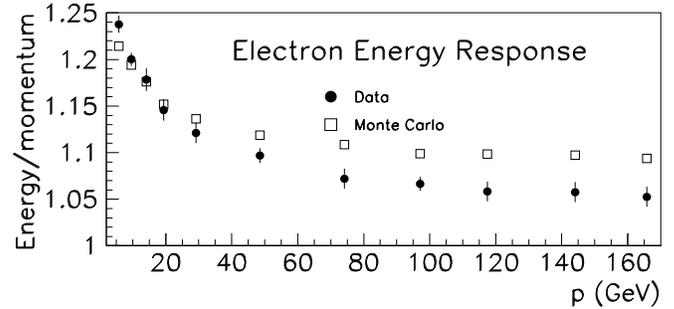,width=.5\textwidth}}
\end{center}
\vspace{-.3in} 
\caption[]{Normalized response of calorimeter to electrons as a function 
of energy for both data (solid circles) and GEANT-based 
Monte Carlo (open squares).}  
\label{fig:elevse} 
\end{figure} 
Figure~\ref{fig:eleeop} shows the shape of the electron energy 
	deposition for 30 and 170~GeV electrons, as well as the fit to the 
	Poisson distribution.  
The detailed GEANT simulation reproduces the 
	shapes of the distributions well, 
	but there are substantial differences 
between the mean values as a function of energy.  

Figure~\ref{fig:elevse} shows the electron energy divided by momentum for 
both data and Monte Carlo.  There are several contributions to the
large non-linearity.  The most important contribution to 
the non-linearity, and the only one present in the GEANT simulation, 
is due to the fact that electrons arrive at the upstream 
edge of the calorimeter and begin showering immediately.  
If one generates the electron showers in the simulation   
equally distributed throughout the steel, as neutrino interactions 
(and hadrons to first order) would be, the GEANT response is 
linear to better than $0.5\%$.  

There are two other effects that are present in the data but not
	in the simulation, and these give rise to the additional 
	non-linearity that is seen in the data.  
The electron response depends critically on the both the muon gain 
	and the hadron/muon gain ratio for the first two counters.  
These ratios are only known to about one per cent, per counter, 
	so this contributes an additional uncertainty 
	which could affect both the scale and the non-linearity.  
Another contribution to the non-linearity is 
	due to the fact that there are cuts on the reconstructed 
	track momentum in the data that cannot be made in the 
	Monte Carlo simulation.  
These cuts do not affect the hadron response because hadrons are 
	extremely unlikely to shower before the calorimeter.  
There is however approximately one radiation length of material 
	distributed throughout the last arm of the calibration 
	beam spectrometer.  
This material is included in the GEANT simulation, but its effect 
	on the tracking efficiency is not.  
This is particularly important at high energies, where the 
	upstream showers are most likely to cause ambiguities in the 
	momentum determination.  
  
One correction to the gains of the first few 
	counters in the detector that is extremely important 
	for measuring the electron response of the calorimeter, is 
	due to the RE effect which is discussed in 
	Section~\ref{sec:eqcor}.

\begin{figure}[tbp] 
\begin{center} 
\centerline{\psfig{figure=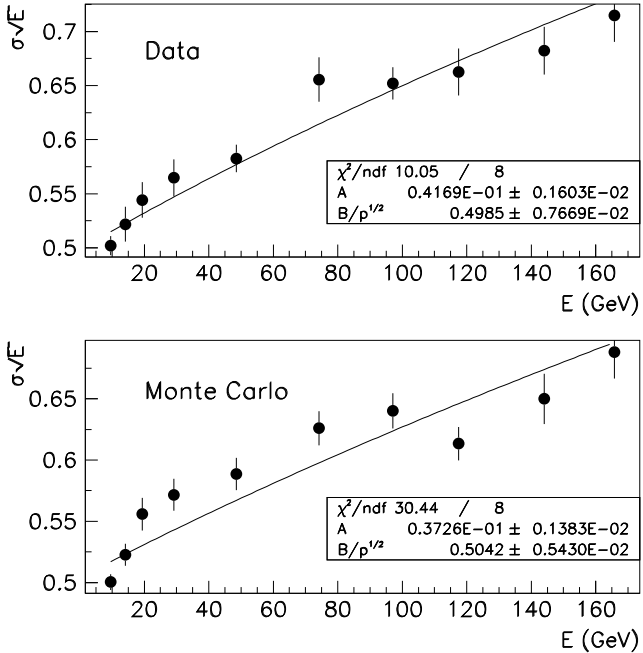,width=.5\textwidth}}
\end{center} 
\vspace{-.3in} 
\caption[]{Electron energy resolution for both data and 
GEANT-based Monte Carlo as a function of energy, and the 
results to a fit of the 
form $\sigma(E)/E = A \oplus \frac{B}{\sqrt{E}}$.} 
\label{fig:elewid} 
\end{figure}

The GEANT-based Monte Carlo predicts the sampling term in the resolution 
	of electron energies 
	to within $2\%$, as can be seen in Figure~\ref{fig:elewid}.  
As with the hadron resolution, the electron resolution can be fit to
	the form 
$\sigma(E)/E = A \oplus \frac{B}{\sqrt{E}} \oplus \frac{C}{E}$, 
and the noise term ($C$) is 
	consistent with zero for both the data and Monte Carlo and is
removed from the fit.  
The sampling term $B$ for electrons is $0.499\pm0.008 (\sqrt{{\rm GeV}})$ from
	data while it is $0.504\pm0.006 (\sqrt{{\rm GeV}})$ in Monte Carlo, 
	showing extremely good agreement. 

Finally, a very important parameter of the calorimeter is the difference
	between electron and hadron responses as a function of energy. 
Section~\ref{sec:hadronmc} describes how the hadron resolution and 
	non-linearity depend critically on this difference.  
In other words, the more similar the electron and hadron responses are, 
	the better the calorimeter resolution, and the more linear.  
The measured electron/hadron difference must be corrected by $1\%$ 
	to account for the fact that the electrons in the calibration beam 
	started upstream, while those from neutrino interactions (or those  
	from hadron showers) are much more uniformly distributed throughout
	each steel plate.  
The ratio of the reconstructed energy of electrons compared to 
	that of hadrons is \hbox {$1.06\pm 0.03$} at 75~GeV, which 
	corresponds to a small non-linearity, in agreement with 
	what is seen in the calorimeter response as a 
	function of energy.

%
\section{Conclusions}
In this paper we outline the calibration technique and subsequent 
	testing of the NuTeV calorimeter.  
The overall gain and time dependence of the calorimeter 
	are tracked using muons from the neutrino beam, 
	and PMT gains are determined to better 
	than one percent for a given run period.  
Using several sets of linear ADC's with different gains, 
	we are able to cover the large dynamic range 
	required to reconstruct minimum energy deposition 
	at the percent level, as well as reconstruct energy 
	deposition from 600~GeV neutrino induced hadron showers.  

Although the technique of calibrating the detector with muons from 
	neutrino interactions may seem simple and straightforward, checking
	this technique requires a very detailed and well-designed calibration 
	beam.  
By using a low mass spectrometer with long lever arm, NuTeV is able to 
	achieve event-by-event momentum resolution of better than 0.1\%.
In addition, as a consequence of careful calibration of the magnets in 
	the spectrometer and measurement of the particle composition 
	of the hadron beam, the absolute hadron energy
	scale of the calorimeter is determined to better than $0.3\%$.  

The non-linearity in the hadron response of the calorimeter is 
	measured and agrees with predictions based on the measured 
	difference between the hadron and electron response at 
	a particular energy. 
Finally, the muon and electron responses of the calorimeter are 
	shown to agree with a GEANT-based Monte Carlo prediction, once the 
	details of the calorimeter geometry are accurately included.  
The vital statistics of the NuTeV calorimeter are summarized in Table 
\ref{tab:vitals}.  
\begin{table}[tbp] 
\caption[]{Vital statistics of the NuTeV calorimeter.} 
\label{tab:vitals}
\begin{center}
\begin{tabular}{|l|l|} 
\hline 
Redeeming Feature & Measurement \\ 
\hline
Hadron Non-linearity & \\
from 5.9~GeV to 190~GeV & $3.0\pm0.5\%$ \\ 
\hline
Hadron Energy Scale Uncertainty & $0.43\%$ \\ 
\hline 
Hadron Energy Resolution: & \\
$\sigma(E)/E = A \oplus \frac{B}{\sqrt{E}} $ & \\ 
A: Constant Term &  $0.022\pm 0.001$ \\ 
B: Stochastic Term $(\sqrt{GeV})$& $0.86\pm 0.01$ \\
\hline 
Residual position dependence of & \\
hadron energy reconstruction & \\ 
more than 50~cm from edge & $<0.5\% $ \\
\hline 
Electron/Hadron Response Ratio & \\
(using Groom Parameterization) & $1.08\pm 0.01$ \\
\hline 
Electron Energy Resolution: & \\ 
A: Constant Term &  $0.042\pm 0.002$ \\ 
B: Stochastic Term $(\sqrt{GeV})$& $0.499\pm 0.008 $ \\
\hline 
Average Number of & \\ 
Photoelectrons/counter/MIP & 30 \\ \hline
Hadron MIP-to-GeV Conversion & \\
Factor ($C_\pi$ GeV/MIP) & $.212\pm .001$ \\ 
\hline
\hline 
\end{tabular} 
\end{center}
\end{table} 
In conclusion, NuTeV has accomplished its goal of calibrating the 
	absolute energy scale of its calorimeter to the level 
	dictated by the physics analyses that NuTeV is performing.  
The calibration beam data also yields a wealth of information 
	about hadron, electron, and muon interactions 
	in an iron-scintillator sampling calorimeter. 
These can be used to study designs of future calorimeters with 
	similar geometries, such as the MINOS detector to 
	search for neutrino oscillations, 
	as well as for space-based calorimeters to measure 
	cosmic ray fluxes~\cite{como}.

\section*{Acknowledgements}
We would like to express our gratitude to the US Department of Energy
    and the US National Science Foundation for their support.     
We express our deep appreciation to Fermilab for providing
     all the necessary technical support in this experiment.
We thank the numerous people at Fermilab and the collaborating institutions
    who helped in all the facets of this project.
At Fermilab, we especially appreciate the help of the Beams Division,
    as well as the Engineering, Mechanical, Electrical, and Survey and 
    Alignment groups.

\end{document}